\documentclass[journal]{journal}
\pdfoutput=1
\usepackage{multicol,lipsum}
\usepackage{booktabs}
\usepackage[normalem]{ulem}
\usepackage{longtable}
\useunder{\uline}{\ul}{}

\ifCLASSINFOpdf
	\usepackage[pdftex]{graphicx}
 
\else
\fi
\usepackage[cmex10]{amsmath}

\usepackage{mathtools}

\usepackage{float}

\usepackage[export]{adjustbox}

\usepackage{fixltx2e}
\usepackage{ulem}
\usepackage{subfigure}
\usepackage{xcolor}
\usepackage{setspace}

\usepackage[colorlinks,
            linkcolor=blue,
            anchorcolor=blue,
            citecolor=blue]{hyperref}

\usepackage{nameref}

\begin{document}
%
\title{Newtonian Mechanics Based Transient Stability PART VI: Machine Transformation}
%
%
%

\author{{Songyan Wang,
        Jilai Yu,  
				Aoife Foley,
        Jingrui Zhang
        }
        
}
%
%

\markboth{Journal of \LaTeX\ Class Files,~Vol.~6, No.~1, January~2007}%
{Shell \MakeLowercase{\textit{et al.}}: Bare Demo of IEEEtran.cls for Journals}
%



\maketitle
\thispagestyle{empty}
\begin{abstract}
This paper focuses on the transformations from the individual machine to the equivalent machine through the ``correction'' perspective of the inner-group machine. The machines are first classified as the real machine with equation of motion and the pseudo machine without equation of motion. Then, it is clarified that both individual machine and equivalent machine are real machines, while the superimposed machine is a pseudo machine. Based on the classifications of the machines, two types of machine transformations are provided.
The two types of machine transformations are based on the ``energy correction'' and ``trajectory correction'' of the inner-group machine, respectively. For the energy correction case, it is clarified that the trajectory transformation completely fails, while the energy transformation mathematically holds yet it is physically meaningless. For the trajectory correction case, it is clarified that both energy transformation and trajectory transformation are established. The reason is that each trajectory-correction based individual machine has the same equation of motion, i.e., the motion of the equivalent Machine-CR. Simulation results show that the machine transformation from the individual machine to the equivalent machine can be realized only through trajectory correction.
\end{abstract}

%

\IEEEpeerreviewmaketitle
\begin{small}

\begin{tabular}{lllll}
    &            &               &                  &                                \\
  \multicolumn{5}{c}{\textbf{Nomenclature}}                                                \\

  DLP    &  \multicolumn{1}{c}{} & \multicolumn{3}{l}{Dynamic liberation point}            \\
  DSP    &                       & \multicolumn{3}{l}{Dynamic stationary point}            \\
  EAC    &                       & \multicolumn{3}{l}{Equal area criterion}                \\
  TSA    &                       & \multicolumn{3}{l}{Transient stability assessment}      \\
  MOD    &                       & \multicolumn{3}{l}{Mode of disturbance}      \\
  COI    &                       & \multicolumn{3}{l}{Center of inertia}      \\
  GTE    &                       & \multicolumn{3}{l}{Global total transient energy}           \\
  IVCS    &                       & \multicolumn{3}{l}{Individual-machine virtual COI-SYS machine}           \\
  LOSP    &                       & \multicolumn{3}{l}{Loss of synchronism point}           \\ 
  IMTR    &                       & \multicolumn{3}{l}{Individual-machine trajectory}           \\ 

\end{tabular}
\end{small}

\section{Introduction} \label{section_I}

%
%
%
%
\subsection{LITERATURE REVIEW} \label{section_IA}
\raggedbottom
\IEEEPARstart{M}{achine} transformation describes the change from the individual machine to the equivalent machine. Following the original definitions, the equivalent machine are depicted as ``motion equivalence'' of the individual machines inside each group \cite{1}-\cite{5}.
However, since both individual machine and equivalent machine strictly follow the machine paradigms and they both show effectiveness in TSA, the equivalent machine analyst attempts to analyze the changes from the individual machine to the equivalent machine from another ``correction'' perspective rather than the original motion equivalence based definitions.
In particular, since the inner group machine motions are the differences between the original system and the equivalent system, the equivalent machine analysts focus on the “removal” of the inner-group machine to realize the transformation from the original system to the equivalent system. This removal can also be expressed as the ``correction of the inner-group machine transient energy'' or the ``correction of the inner-group machine trajectory''.
\par The machine transformation can be seen as a most mysterious conjecture in the history of power system transient stability. The reason is that the machine transformation comprises of the complicated relationship among individual machine, equivalent machine, inner-group machine and the superimposed machine, although it only focuses on the transformation from the individual machine to the equivalent machine. The related studies are rare and most conjectures remain unsolved even nowadays. In the early history of the transient stability analysis, Fouad conjectured that ``not all the excess kinetic energy at fault clearing contributes directly to the separation of the system, and part of kinetic energy should be corrected'' \cite{6}, \cite{7}.
This conjecture also supports modern equivalent machine theory. Xue \cite{8} and Fang \cite{9} focus on the correction of the inner-group machine transient energy. They found that the both KE and PE corrections may fully ensure the strict NEC (EAC) characteristic in the equivalent machine. The success of NEC in the equivalent machine also inspires the equivalent machine analysts to believe that it is the ``energy correction'' that supports the machine transformations. However, this energy correction leads to the emergence of two questions: 
\\(i) Could transient energy be subtracted or added in the transient stability analysis?
\\(ii) Could machine transformation be realized from another perspective rather than the energy correction of the inner-group machine?
\par The exploration of the energy correction may theoretically reveal the physical nature of the transient energy. In addition, it may also help readers take a deep insight into the relationship between the individual machine and the equivalent machine from another distinctive ``trajectory correction'' perspective.

\subsection{SCOPE AND CONTRIBUTION OF THE PAPER} \label{section_IB}
In this paper, the transformations from the individual machine to the equivalent machine by using ``energy correction'' and ``trajectory correction'' are analyzed.
Through the revisit of the Newtonian system, the balls in the Newtonian system are classified as the real ball with equation of motion and the pseudo balls without equation of motion. Then, the classification of balls in the Newtonian system is extended to the power system transient stability analysis. It is clarified that both individual machine and equivalent machine are real machines with equation of motions, while the superimposed machine is a pseudo machine without equation of motion.
Based on the classification of machines, two types of machine transformations, i.e., the ``energy correction'' based machine transformation and the ``trajectory correction'' based machine transformation are provided. For the energy correction case, it is clarified that the inner-group transient energy correction based individual machine (ECIM) is a pseudo machine without equation of motion, and thus the trajectory transformation completely fails.
In addition, although the superimposition of the ECIMTEs is equal to EMTE, this energy superimposition is a ``mathematical coincidence'' because it is physically meaningless.
For the trajectory correction case, although the trajectory correction based individual machine (TCIM) is also a pseudo machine, the motion of each TCIM is set the same with that of Machine-CR. Against this background, both trajectory transformation and the energy transformation are established in the TCIM. The analysis in this paper clarifies that the machine transformation can be realized only through trajectory correction.
\par  The contributions of this paper are summarized as follows:
\\ (i) The classifications of the machines in the power system transient stability are given. This machine classification emphasizes the dominant role of the equation of motion of the machine in TSA.
\\ (ii) Two types of machine transformations are provided. The two transformations provide a deep insight into the relationship between the individual machine and equivalent machine.
\\ (iii) The problems of the inner-group machine transient energy correction is analyzed. This clarifies that the machine transformation can be realized only through trajectory corrections.
\par The reminder of the paper is organized as follows. In Section \ref{section_II}, the classifications of the balls in the Newtonian system are provided. In Section \ref{section_III}, classifications of machines in the power system transient stability are given. In Section \ref{section_IV}, the machine transformation through original definition (motion equivalence) is revisited.
In Section \ref{section_V}, the machine transformation through energy correction is provided. In Section \ref{section_VI}, the machine transformation through trajectory correction is analyzed. In Section \ref{section_VII}, simulation results show the failure and success of the energy correction and trajectory correction, respectively. Conclusions are given in Section \ref{section_VIII}.
\par In this paper, the analysis about the equivalent CR-NCR system is based on its mirror system, i.e., the CR-SYS system \cite{4}. That is, all the analysis in this paper about the equivalent machine is based on the CR-SYS system. The deductions in the CR-SYS system can be naturally extended to the situations in NCR-SYS system.
\par The authors strongly emphasize that the individual machine and the equivalent machine are ``companions'' in TSA as analyzed in Ref. \cite{5}. The analysis of machine transformation in this paper just provides a deep insight into the relationship between the individual machine and equivalent machine.

\section{CLASSIFICATION OF BALLS IN THE NEWTONIAN SYSTEM} \label{section_II}
\subsection{REAL BALL WITH EQUATION OF MOTION}  \label{section_IIA}

In this section, the classification of balls is first analyzed through the visible Newtonian system. Then, it is extended to the classifications of machines in the power system transient stability.
\par For the Newtonian system that is formed by the ball and Earth, the motion of each ball is depicted as
\begin{equation}
  \label{equ1}
  \left\{\begin{array} { l } 
    { \frac { d h _ {\mathrm { ball } } } { d t } = v _ {\mathrm { ball } } }  \\
    \\
    { m _ {\mathrm { ball } } \frac { d v _ { \mathrm { ball } } } { d t } = F _ {\mathrm { ball } } }
    \end{array} \quad \left\{\begin{array}{l}
        \frac { d h _ {\mathrm { Eath } } } { d t } = v _ {\mathrm { Eath } } \\
      \\
    m_{\mathrm {Earth }} \frac{d v_{\mathrm {Eath }}}{d t}=F_{\mathrm{Earth }}
    \end{array}\right.\right.
\end{equation}
\par
\begin{figure}[H]
  \centering
  \includegraphics[width=0.3\textwidth,center]{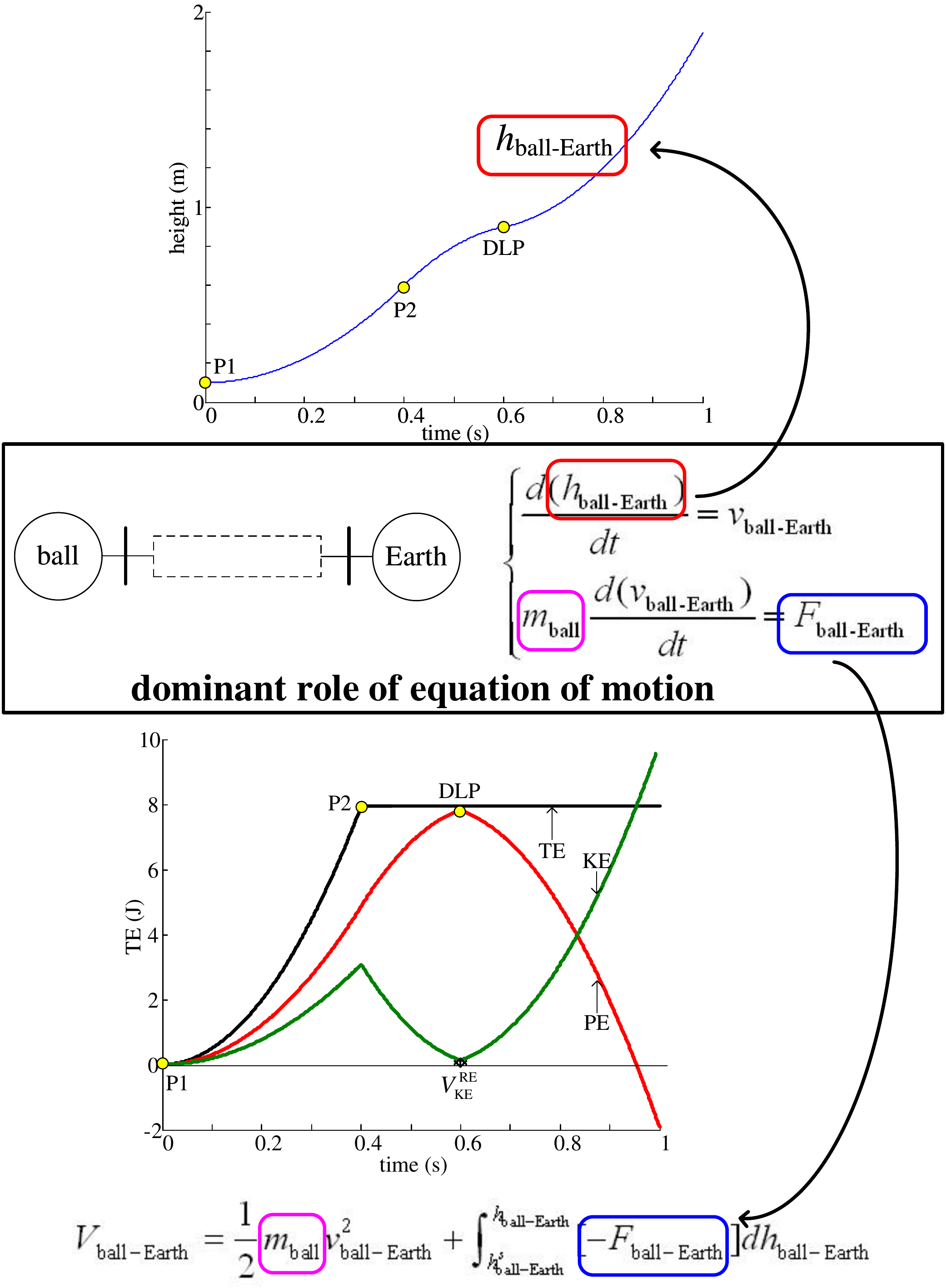}
  \caption{Dominant role of the relative equation of motion.} 
  \label{fig1}  
\end{figure}

 In Eq. (\ref{equ1}), \textit{F} and \textit{h} are the net force and the altitude of each ball, respectively. Other parameters are shown in Fig. \ref{fig1}.
\par Following Eq. (\ref{equ1}), the relative motion between the ball and Earth can be given as
\begin{equation}
  \label{equ2}
  \left\{\begin{array}{l}
    \frac{d\left(h_{\mathrm{ball}\mbox{-}\mathrm{Earth}}\right)}{d t}=v_{\mathrm{ball}\mbox{-}\mathrm{Earth}} \\
    \\
    m_{\mathrm {ball }} \frac{d\left(v_{\mathrm{ball}\mbox{-}\mathrm{Earth}}\right)}{d t}=F_{\mathrm{ball}\mbox{-}\mathrm{Earth}}
    \end{array}\right.
\end{equation}
where
\begin{spacing}{1.5}
  \noindent$h_{\mathrm{ball}\mbox{-}\mathrm{Earth}}=h_{\mathrm{ball}}-h_{\mathrm{Earth}}$\\
  $v_{\mathrm{ball}\mbox{-}\mathrm{Earth}}=v_{\mathrm{ball}}-v_{\mathrm{Earth}}$\\
  $F_{\mathrm{ball}\mbox{-}\mathrm{Earth}}=F_{\mathrm{ball}}-\frac{m_{\mathrm{ball}}}{m_{\mathrm{Earth}}}F_{\mathrm{Earth}}$
\end{spacing}
In this paper, the ball that is modeled based on the relative equation of motion as given in Eq. (\ref{equ3}) is named the ``real ball''. the relative equation of motion of the real ball with respect to Earth has following characteristics.
\vspace*{0.5em}
\\ (i) The relative equation of motion of the real ball should be defined under a pre-given motion reference (Earth).
\\ (ii) The mass of the ball ($m_{\mathrm{ball}}$) is preserved in the equation of motion of the real ball.
\\ (iii) The effect of the Earth to the real ball is reflected through $-m_{\mathrm{ball}/m_{\mathrm{Earth}}G_{\mathrm{Earth}}}$.
\vspace*{0.5em}
\par The relative equation of motion of the real ball with respect to Earth is of key importance in the Newtonian system, because both the trajectory and the energy of the ball are obtained through this equation of motion. The mechanism of the relative equation of motion is shown in Fig. \ref{fig1}.
\vspace*{-0.5em}
Following the equation of motion of the real ball, the ``real'' Newtonian energy is defined as 
\begin{equation}
  \label{equ3}
    V_{\mathrm{ball}}=\frac{1}{2}m_{\mathrm{ball}}v_{\mathrm{ball}\mbox{-}\mathrm{Earth}}^2+
    \int_{h_{\mathrm {ball }}^{s}}^{h_{\mathrm {ball }}}\left[-F_{\mathrm {ball}\mbox{-}\mathrm{Earth} }\right] d h_{\mathrm {ball}\mbox{-}\mathrm{Earth}}
\end{equation}
\par In Eq. (\ref{equ3}), the mass in the Newtonian energy is defined by the mass of the real ball ($m_{\mathrm{ball}}$).
\par Generally, the Earth is set as stationary in the Newtonian system ($m_{\mathrm{Earth}}=\infty$,  $h_{\mathrm{Earth}}=0$, $v_{\mathrm{Earth}}=0$). Then, Eq. (\ref{equ3}) can be further simplified as\
\begin{equation}
  \label{equ4}
  V_{\mathrm{ball}}=\frac{1}{2}m_{\mathrm{ball}}v_{\mathrm{ball}}^2+
  \int_{h_{\mathrm {ball }}^{s}}^{h_{\mathrm {ball }}}\left[-F_{\mathrm {ball}}\right] d h_{\mathrm {ball}}
\end{equation}
\par Following the analysis in Ref. \cite{8}, for a generalized Newtonian system that is formed by multiple balls, the motion of each ball is unique and different because $F_{\mathrm{ball}}/m_{\mathrm{ball}}$, i.e., the acceleration of each ball is different. Against this background, in the generalized Newtonian system, the following deductions can be obtained
\vspace*{0.5em}
\\ (i) The trajectory of each ball is unique and different.
\\ (ii) The NEC inside each ball is also unique and different.
\vspace*{0.5em}
\par In fact, both (i) and (ii) hold because of the equation of motion of each ball is unique and different. (i) and (ii) further indicates the following
\vspace*{0.5em}
\par \textit{The energy computation among balls is meaningless in the Newtonian system if these balls lie in different gravitational fields}.
\vspace*{0.5em}
\par The deduction above can be expressed in a contrary manner.
\vspace*{0.5em}
\par \textit{The energy computation among balls will become meaningful if all balls lie in the same gravitational field}.
\vspace*{0.5em}
\par The deduction above is the well-known ``big ball-small-ball'' Galileo game.
\par
\begin{figure}[H]
  \centering
  \includegraphics[width=0.45\textwidth,center]{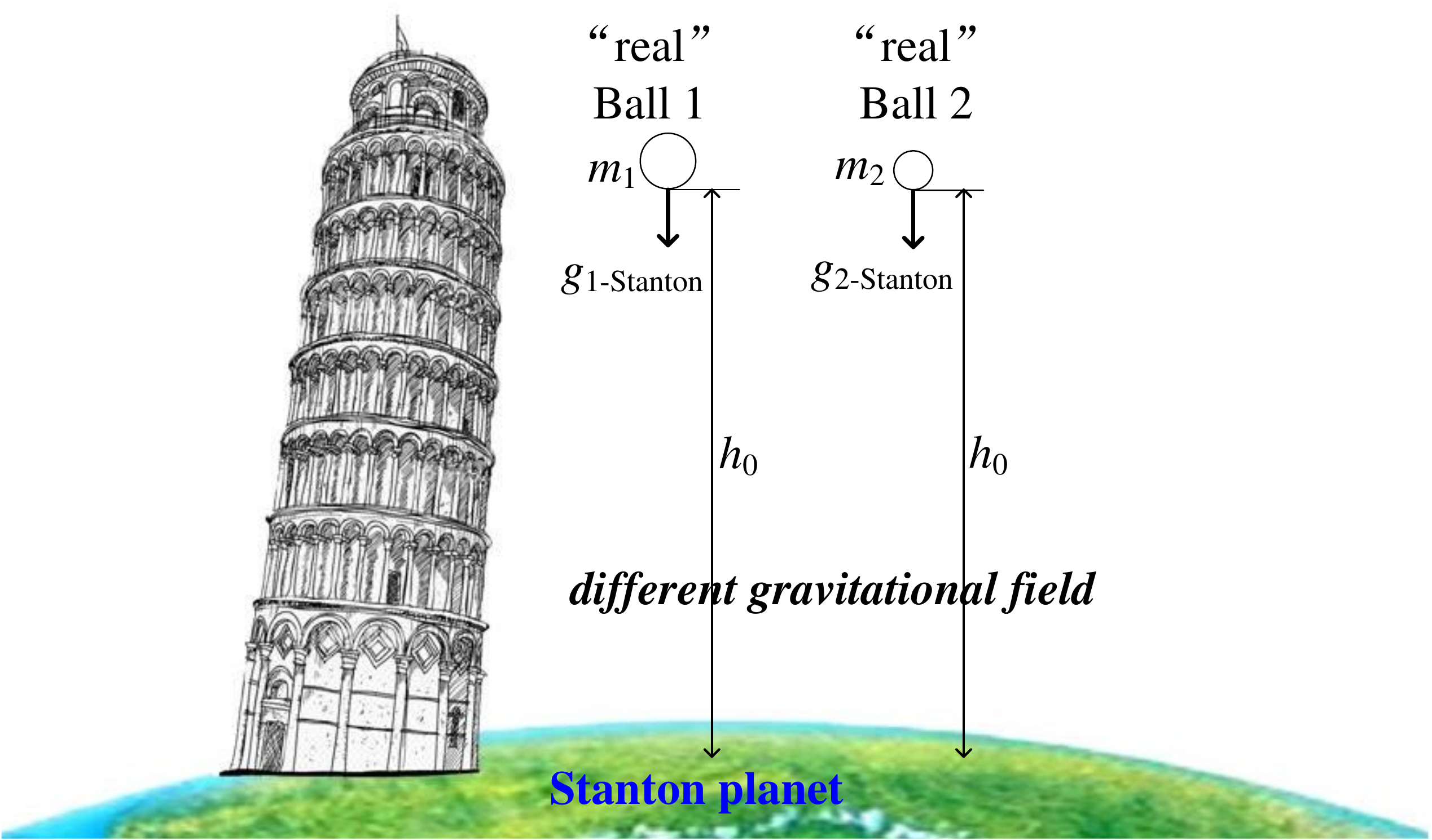}
  \caption{Mechanism of the real ball.} 
  \label{fig2}  
\end{figure}

 The meaningless energy computation in the Newtonian system is demonstrated as below. A simple Newtonian system is shown in Fig. \ref{fig2}.
This Newtonian system is formed by two real balls and Earth. The parameters are shown in Table \ref{table1}. at $t_0$, the two balls remain stationary at the same height. Note that the gravitational field of each ball is different.
\vspace*{-1em}
\begin{table}[H]
  \centering
  \caption{Parameters of the original Newtonian system}
  \label{table1}
  \begin{tabular}{p{0.5cm}cp{0.5cm}cp{0.5cm}}
  \toprule
  &Parameters&    & Value                   &  \\ \midrule
  &$m_1$     &    & 1.0 kg                  &  \\ \midrule
  &$m_2$     &    & 0.5 kg                  &   \\
  &$g_1$     &    & 9.8 $\text{m/s}^2$      &    \\
  &$g_2$     &    & 6.0 $\text{m/s}^2$       &           \\
  &$h_0$     &    & 8.0 m                     &     \\ \bottomrule
  \end{tabular}
\end{table}
\vspace*{-0.5em}
The trajectory of each real ball is depicted as
\begin{equation}
  \label{equ5}
  \left\{\begin{array} { l } 
    { v _ { 1 } = g _ { 1 } t } \\
    \\
    { h _ { 1 } = h _ { 0 } - \frac { 1 } { 2 } g _ { 1 } t ^ { 2 } }
    \end{array} \quad \left\{\begin{array}{l}
    v_{2}=g_{2} t \\
    \\
    h_{2}=h_{0}-\frac{1}{2} g_{2} t^{2}
    \end{array}\right.\right.
\end{equation}
\par In Eq. (\ref{equ5}), note that $g_1$ and $g_2$ are different because the two balls stay on the Stanton Planet \cite{8}.
\par The Newtonian energy of each real ball is given as
\begin{equation}
  \label{equ6}
  \left\{\begin{array}{l}
    V_{1}=\frac{1}{2} m_{1} v_{1}^{2}+m_{1} g_{1} h_{1} \\
    \\
    V_{i}=\frac{1}{2} m_{2} v_{2}^{2}+m_{2} g_{2} h_{2}
    \end{array}\right.
\end{equation}
\par
\begin{figure}[H]
  \centering
  \includegraphics[width=0.42\textwidth,center]{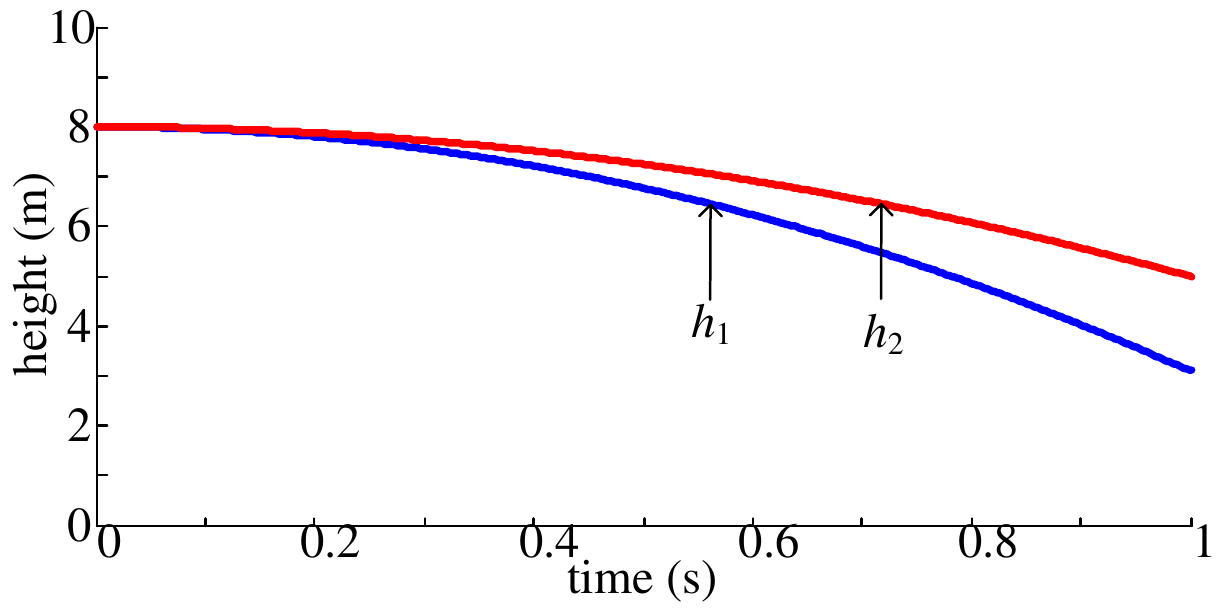}
  \caption{Trajectory of each real ball.} 
  \label{fig3}  
\end{figure}

 The trajectory of each ball is shown in Fig. \ref{fig3}. The NEC inside each ball is shown in Figs. \ref{fig4} (a) and (b), respectively.
\vspace*{-1em}
\begin{figure} [H]
  \centering 
  \subfigure[]{%
  \label{fig4a}
    \includegraphics[width=0.37\textwidth]{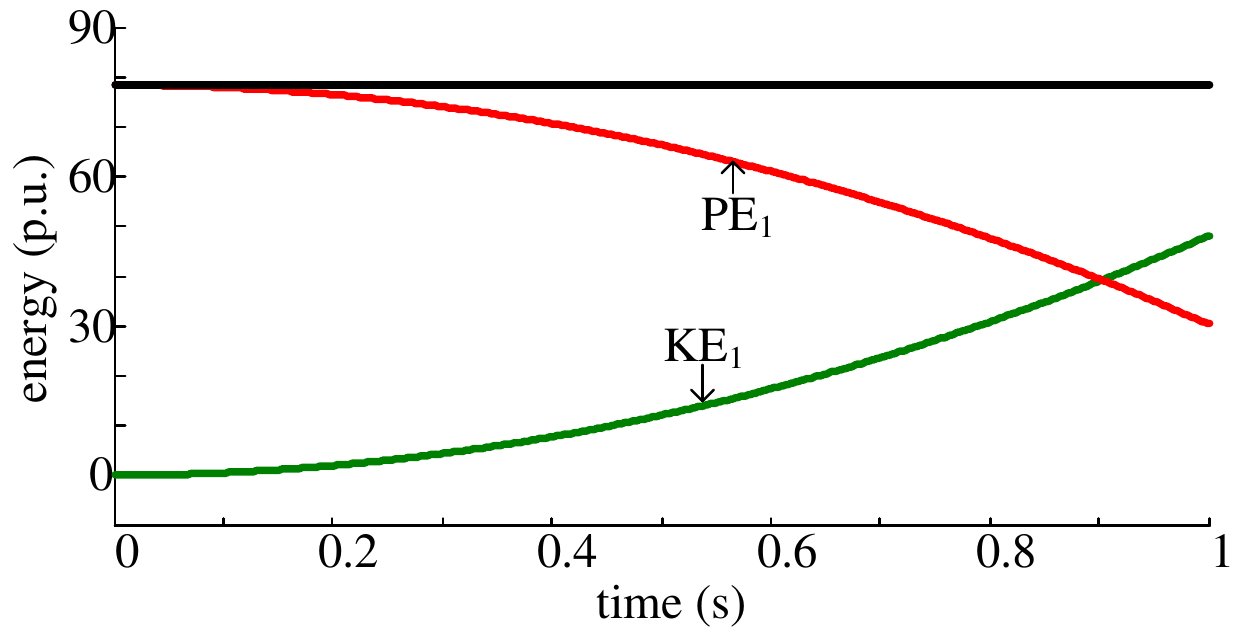}}%
\end{figure} 
\vspace*{-2em}
\addtocounter{figure}{-1}       
\begin{figure} [H]
  \addtocounter{figure}{1}      
  \centering 
  \subfigure[]{%
    \label{fig4b}
    \includegraphics[width=0.37\textwidth]{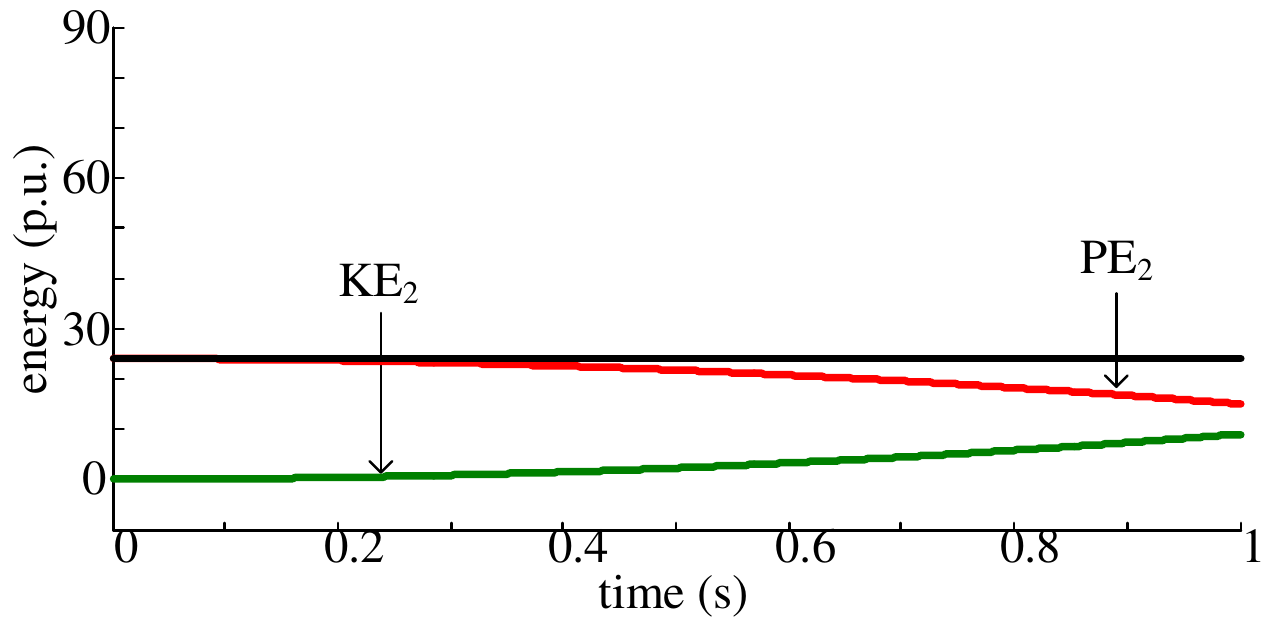}}%
  \caption{NEC inside each real ball. (a) Ball 1. (b) Ball 2.}%
  \label{fig4}
\end{figure}
\vspace*{-0.5em}
From Figs. \ref{fig3} and \ref{fig4}, because the gravitational field of each ball is different, the trajectory ($h_i$) and NEC inside each ball are also unique and different.

\subsection{PSEUDO BALL WITHOUT EQUATION OF MOTION} \label{section_IIB}
The analysis in Section \ref{section_IIA} emerges one question: what might happen if the energy among different balls are added or subtracted?
\par In fact, the addition or subtraction of the NEC will create a ``pseudo'' ball. The energy conversion of the ball is depicted as
\begin{equation}
  \label{equ7}
  V_{\mathrm{pseudo}}=V_{KE\mathrm{pseudo}}+V_{PE\mathrm{pseudo}}
\end{equation}
\par From Eq. (\ref{equ7}), the pseudo ball only has energy conversion. This energy conversion does not have its corresponding equation of motion. That is, the pseudo ball does not have trajectory, mass and force on it. In other words, it does not exist in the physically real Newtonian system. This indicates the following
\vspace*{0.5em}
\par \textit{The pseudo ball is meaningless for the stability analysis of the Newtonian system if it does not have equation of motion}.
\vspace*{0.5em}
\par In brief, the pseudo ball is a mathematical concept that is created through the energy computation among different balls in the Newtonian system.
\par
\begin{figure}[H]
  \centering
  \includegraphics[width=0.42\textwidth,center]{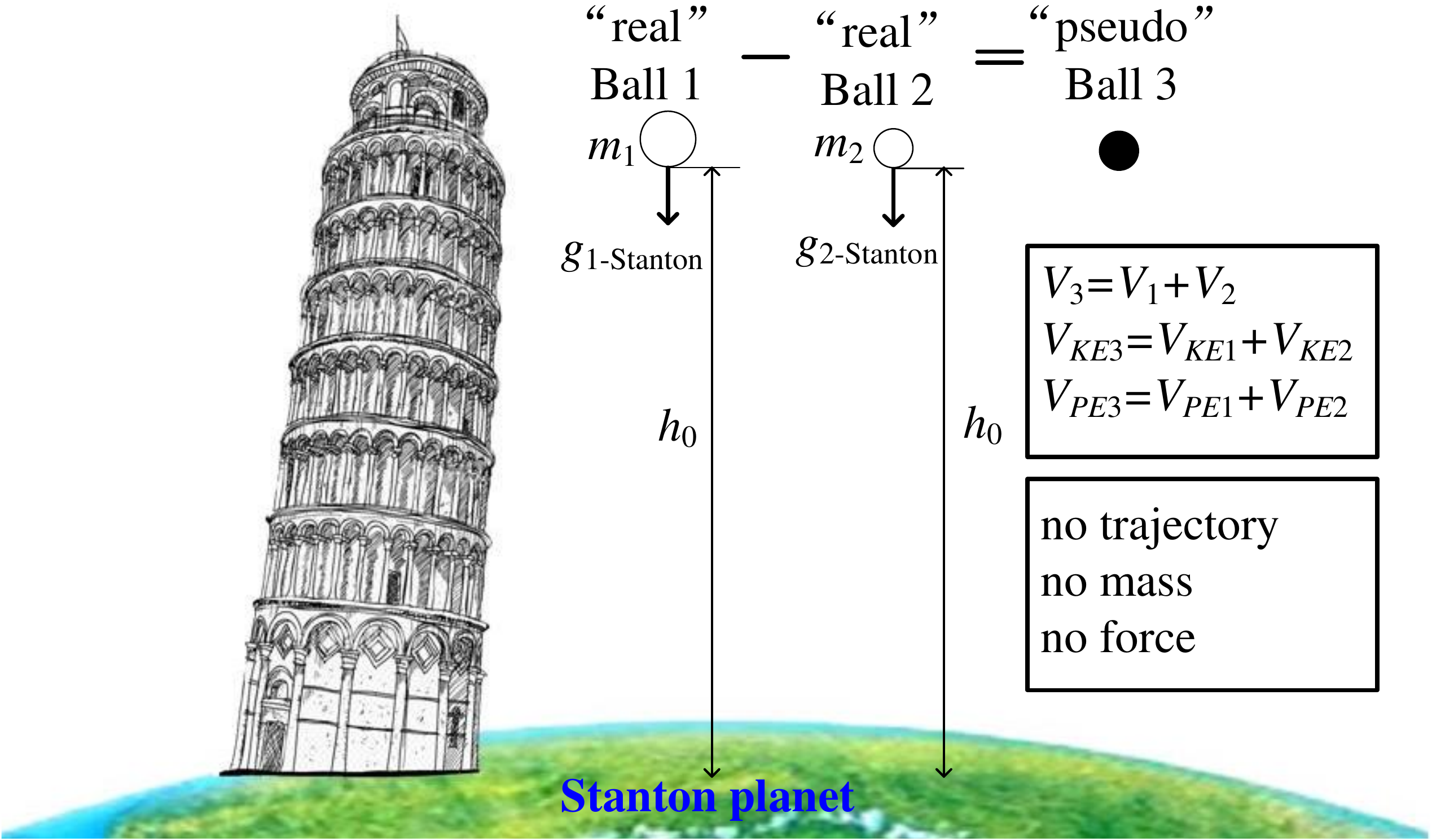}
  \caption{Mechanism of the pseudo ball.} 
  \label{fig5}  
\end{figure}

 A tutorial example about the pseudo ball is given below. The pseudo Ball-3 is fictionally created through the subtraction of the NEC of Ball-1 from that of Ball-2. The pseudo Ball-3 is shown in Fig. \ref{fig5}.
\vspace*{-0.5em}
The energy of pseudo Ball-3 is defined as
\begin{equation}
  \label{equ8}
  V_3=V_1-V_2
\end{equation}
where
\begin{spacing}{1.5}
  \noindent$V_{K E 3}=V_{K E 1}-V_{K E 2}=\frac{1}{2} m_{1} v_{1}^{2}-\frac{1}{2} m_{2} v_{2}^{2}$\\
  $V_{P E 3}=V_{P E 1}-V_{P E 2}=m_{1}g_{1}h_{1}-m_{2}g_{2}h_{2}$
\end{spacing}
\par The energy conversion of pseudo Ball-3 is shown in Fig. \ref{fig6}.
\begin{figure}[H]
  \centering
  \includegraphics[width=0.42\textwidth,center]{fig_6.pdf}
  \caption{NEC inside pseudo Ball 3.} 
  \label{fig6}  
\end{figure}
\vspace*{-0.5em}
From Fig. \ref{fig6}, although the energy conversion inside pseudo Ball-3 is similar to that of real ball as shown in Fig. \ref{fig4}, this energy conversion is completely pseudo because it is simply created through the subtraction between the energy of the two balls. The pseudo Ball 3 does not have any equation of motion. It is a fictional mathematical concept without mass, trajectory and force on it. Clearly, it is meaningless to use pseudo Ball 3 in the stability analysis of the Newtonian system.

\subsection{ADVANTAGES OF THE REAL BALL} \label{section_IIC}
Based on the analysis in Sections \ref{section_IIA} to \ref{section_IIB}, the ``real'' ball can be seen as the fundamental element of the Newtonian system. This is because both the ``real'' trajectory and the ``real'' energy conversion are modeled based on the equation of motion of the real ball. In other, words, the physical nature of the Newtonian system can be reflected only through real ball.
\par Comparatively, the pseudo ball is fictionally created through the energy computation among different balls. The pseudo ball does not have any equation of motion. In particular, it does not have corresponding trajectory, mass and force on it. Therefore, the pseudo ball is meaningless in the stability analysis of the Newtonian system. The physical nature of the Newtonian system is unable to be reflected through the real ball.
\par In the following section, the classifications of balls in the Newtonian system will be extended to the classifications of machines in a multi-machine power system.

\section{REAL INDIVIDUAL MACHINE AND REAL EQUIVALENT MACHINE} \label{section_III}
\subsection{DEFINITION OF REAL MACHINE} \label{section_IIIA}
Following the analysis in Section \ref{section_IIA}, using RM as the motion reference, the equation motion of the real machine is given as
\begin{equation}
  \label{equ9}
  \left\{\begin{array}{l}
    \frac{d \delta_{i\mbox{-}\mathrm{RM}}}{d t}=\omega_{i\mbox{-}\mathrm{RM}} \\
    \\
    M_{i} \frac{d \omega_{i\mbox{-}\mathrm{RM}}}{d t}=f_{i\mbox{-}\mathrm{RM}}
    \end{array}\right.
\end{equation}
where
\begin{spacing}{1.5}
  \noindent$\delta_{i\mbox{-}\mathrm{RM}}=\delta_{i}-\delta_{\mathrm{RM}}$\\
  $\omega_{i\mbox{-}\mathrm{RM}}=\omega_{i}-\omega_{\mathrm{RM}}$\\
  $f_{i\mbox{-}\mathrm{RM}}=P_{i}-\frac{M_i}{M_{\mathrm{RM}}}P_{\mathrm{RM}}$
\end{spacing}
Following the mechanisms of the real balls, the real machine in the RM reference shows the following characteristics
\vspace*{0.5em}
\\ (i) The relative equation of motion is defined under RM reference.
\\ (ii) The inertia of the machine ($M_i$) is preserved in the equation.
\\ (iii) The effect of the RM to the real machine is reflected through $-M_{i}/M_{\mathrm{RM}}P_{\mathrm{RM}}$.
\vspace*{0.5em}
\par The real machine transient energy is defined as
\begin{equation}
  \label{equ10}
  V_{i\mbox{-}\mathrm{RM}}=\frac{1}{2} M_{i} \omega_{i\mbox{-}\mathrm{RM}}^{2}+\int_{\delta_{i\mbox{-}\mathrm{RM}}^{s}}^{\delta_{i\mbox{-}\mathrm{RM}}}\left[-f_{i\mbox{-}\mathrm{RM}}^{(P F)}\right] d \delta_{i\mbox{-}\mathrm{RM}}
\end{equation}
\par
\begin{figure}[H]
  \centering
  \includegraphics[width=0.42\textwidth,center]{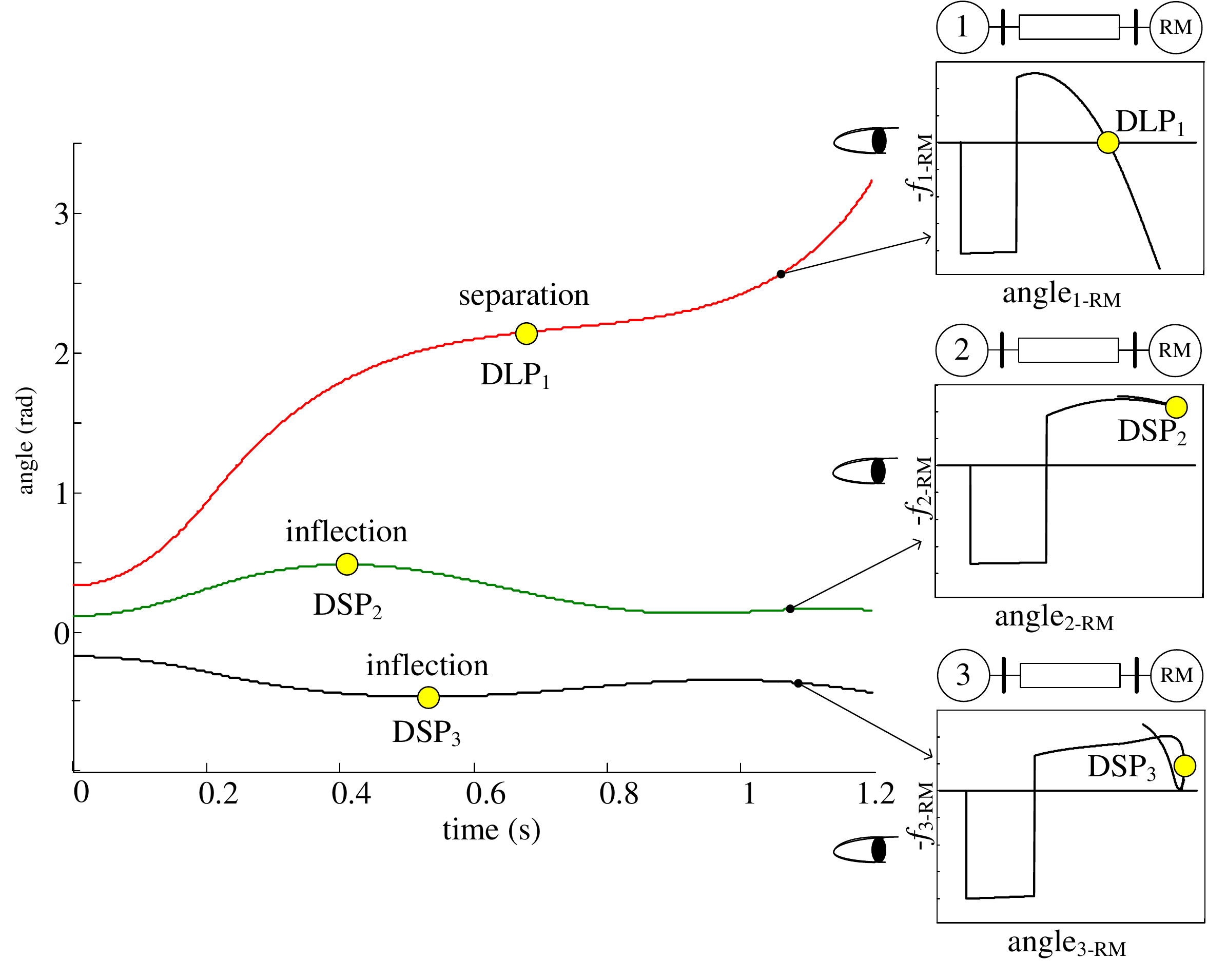}
  \caption{Mechanisms of the real machine.} 
  \label{fig7}  
\end{figure}

\begin{figure}[H]
  \centering
  \includegraphics[width=0.42\textwidth,center]{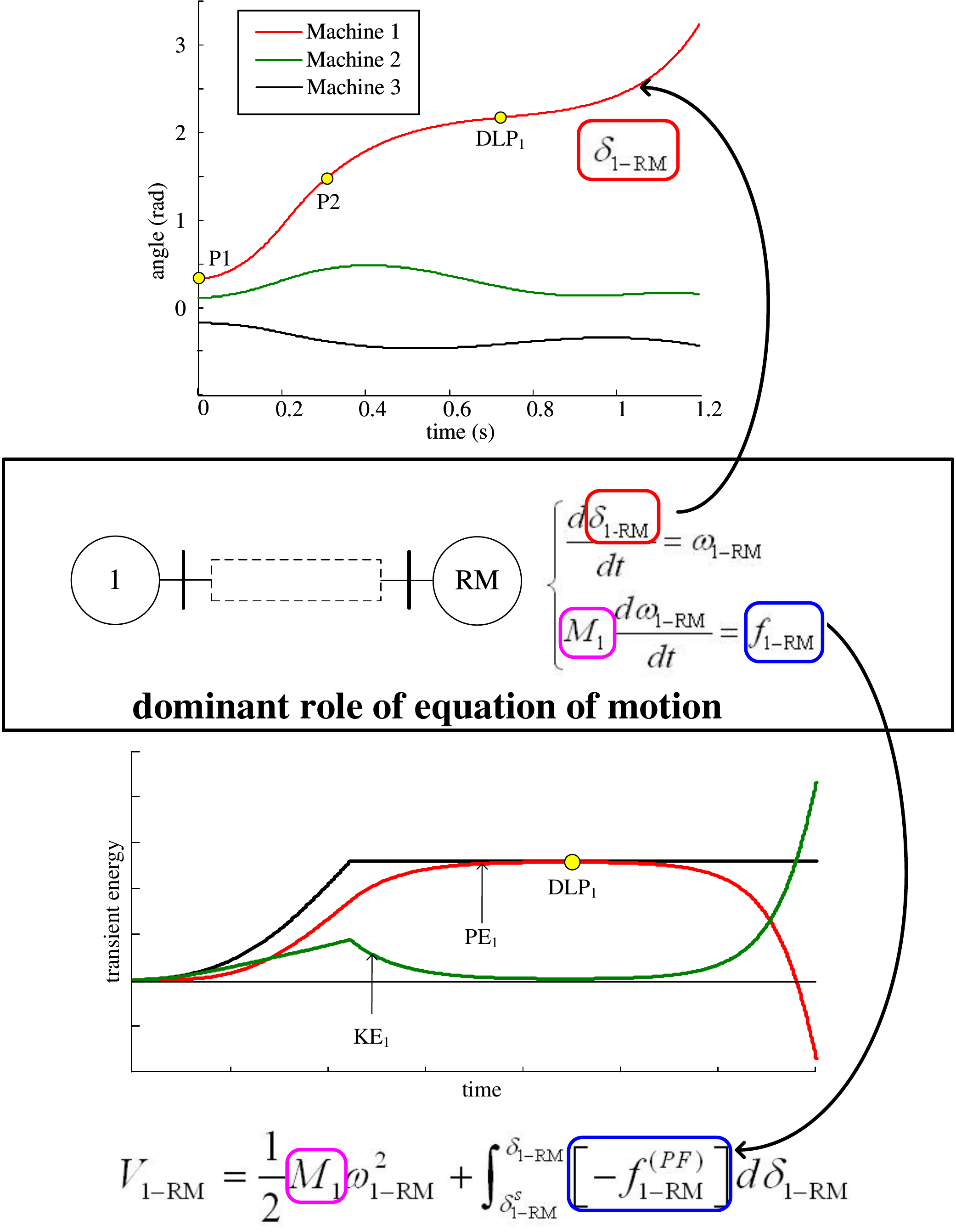}
  \caption{Correlation between trajectory and energy in the real machine.} 
  \label{fig8}  
\end{figure}

 From analysis above, in brief, the key characteristic of the real machine is that it is modeled based on the equation of motion with its corresponding inertia ($M_i$). The mechanism of the real machine is shown in Fig. \ref{fig7}.
\vspace*{-0.5em}
From Fig. \ref{fig7}, strict correlation between trajectory and energy of the real machine is established through the relative equation of motion of the real machine. The machine paradigms are defined based on the real machine \cite{1}. The correlation between trajectory and energy in the real machine is shown in Fig. \ref{fig8} \cite{1}.
\vspace*{-0.5em}
Compared with real machine that has its corresponding equation of motion, the pseudo machine is the machine that is created from the energy computation without any equation of motion, or its original trajectory is replaced by other machines.
\par Following the analysis in Refs. \cite{2}, \cite{4}, both individual machine and equivalent machine are real machines, while the superimposed machine is a typical pseudo machine. Detailed analysis is given in the following sections. 

\subsection{CLASSIFICATIONS OF MACHINES} \label{section_IIIB}
\vspace*{0.5em}
\noindent 1) ``REAL'' INDIVIDUAL MACHINE
\vspace*{0.5em}
\par The individual machine is a typical real machine \cite{2}.
\par The relative motion of Machine \textit{i} in the COI-SYS reference is given as
\begin{equation}
  \label{equ11}
  \left\{\begin{array}{l}
    \frac{d \delta_{i\mbox{-}\mathrm{SYS}}}{d t}=\omega_{i\mbox{-}\mathrm{SYS}} \\
    \\
    M_{i} \frac{d \omega_{i\mbox{-}\mathrm{SYS}}}{dt}=f_{i\mbox{-}\mathrm{SYS}}
    \end{array}\right.
\end{equation}
\par In Eq. (\ref{equ11}), the individual machine is a real machine with its physical inertia $M_i$. Detailed parameters were already given in Ref. \cite{2}.
\par  The IMTE is defined in a typical Newtonian energy manner. The IMTE is defined as
\begin{equation}
  \label{equ12}
  V_{i\mbox{-}\mathrm{SYS}}=V_{K E i\mbox{-}\mathrm{SYS}}+V_{P E i\mbox{-}\mathrm{SYS}}
\end{equation}
where
\begin{spacing}{2}
  \noindent$V_{K E i\mbox{-}\mathrm{SYS}}=\frac{1}{2} M_{i} \omega_{i\mbox{-}\mathrm{SYS}}^{2}$\\
  $V_{P E i\mbox{-}\mathrm{SYS}}=\int_{\delta_{i\mbox{-}\mathrm{SYS}}^{s}}^{\delta_{i\mbox{-}\mathrm{SYS}}}\left[-f_{i\mbox{-}\mathrm{SYS}}^{(P F)}\right] d \delta_{i\mbox{-}\mathrm{SYS}}$
\end{spacing}
Detailed parameters were already given in Ref. \cite{2}.
\par
\begin{figure}[H]
  \centering
  \includegraphics[width=0.42\textwidth,center]{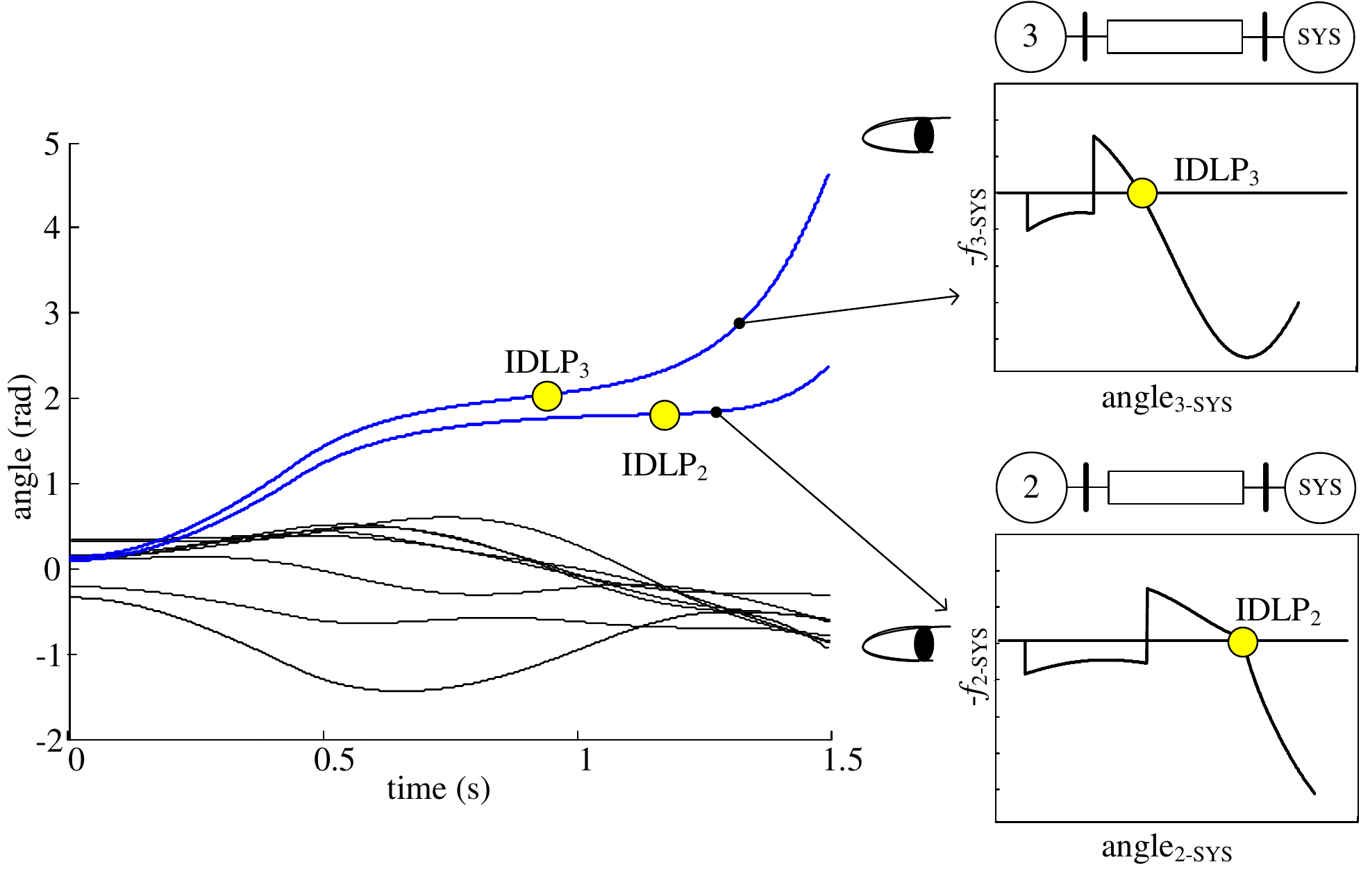}
  \caption{Mechanisms of the real individual machine [TS-1, bus-4, 0.447 s].} 
  \label{fig9}  
\end{figure}

 The mechanism of the individual machine is shown in Fig. \ref{fig9}. From the figure, strict correlation between each IMTR and IMTE is established through equation of motion of the individual machine.
\vspace*{-0.5em}
\noindent 2) ``REAL'' EQUIVALENT MACHINE
\vspace*{0.5em}
\par The equivalent Machine-CR is also a typical real machine \cite{4}.
\begin{equation}
  \label{equ13}
  \left\{\begin{array}{l}
    \frac{d \delta_{\mathrm{CR}\mbox{-}\mathrm{SYS}}}{d t}=\omega_{\mathrm{CR}\mbox{-}\mathrm{SYS}} \\
    \\
    M_{\mathrm{CR}} \frac{d \omega_{\mathrm{CR}\mbox{-}\mathrm{SYS}}}{d t}=f_{\mathrm{CR}\mbox{-}\mathrm{SYS}}
    \end{array}\right.
\end{equation}
\par In Eq. (\ref{equ13}), Machine-CR is a real machine with its equivalent inertia $M_{\text{CR}}$. Detailed parameters were already given in Ref. \cite{4}.
\par The EMTE is defined in a typical Newtonian energy manner. The EMTE is defined as
\begin{equation}
  \label{equ14}
  V_{\mathrm{CR}\mbox{-}\mathrm{SYS}}=V_{K E \mathrm{CR}\mbox{-}\mathrm{SYS}}+V_{P E \mathrm{CR}\mbox{-}\mathrm{SYS}}
\end{equation}
where
\begin{spacing}{2}
  \noindent$V_{K E \mathrm{CR}\mbox{-}\mathrm{SYS}}=\frac{1}{2} M_{\mathrm{CR}} \omega_{\mathrm{CR}\mbox{-}\mathrm{SYS}}^{2}$\\
  $V_{P E \mathrm{CR}\mbox{-}\mathrm{SYS}}=\int_{\delta_{\mathrm{CR}\mbox{-}\mathrm{SYS}}^{s}}^{\delta_{\mathrm{CR}\mbox{-}\mathrm{SYS}}}\left[-f_{\mathrm{CR}\mbox{-}\mathrm{SYS}}^{(P F)}\right] d \delta_{\mathrm{CR}\mbox{-}\mathrm{SYS}}$
\end{spacing}
Detailed parameters were already given in Ref. \cite{4}.
\par
\begin{figure}[H]
  \centering
  \includegraphics[width=0.45\textwidth,center]{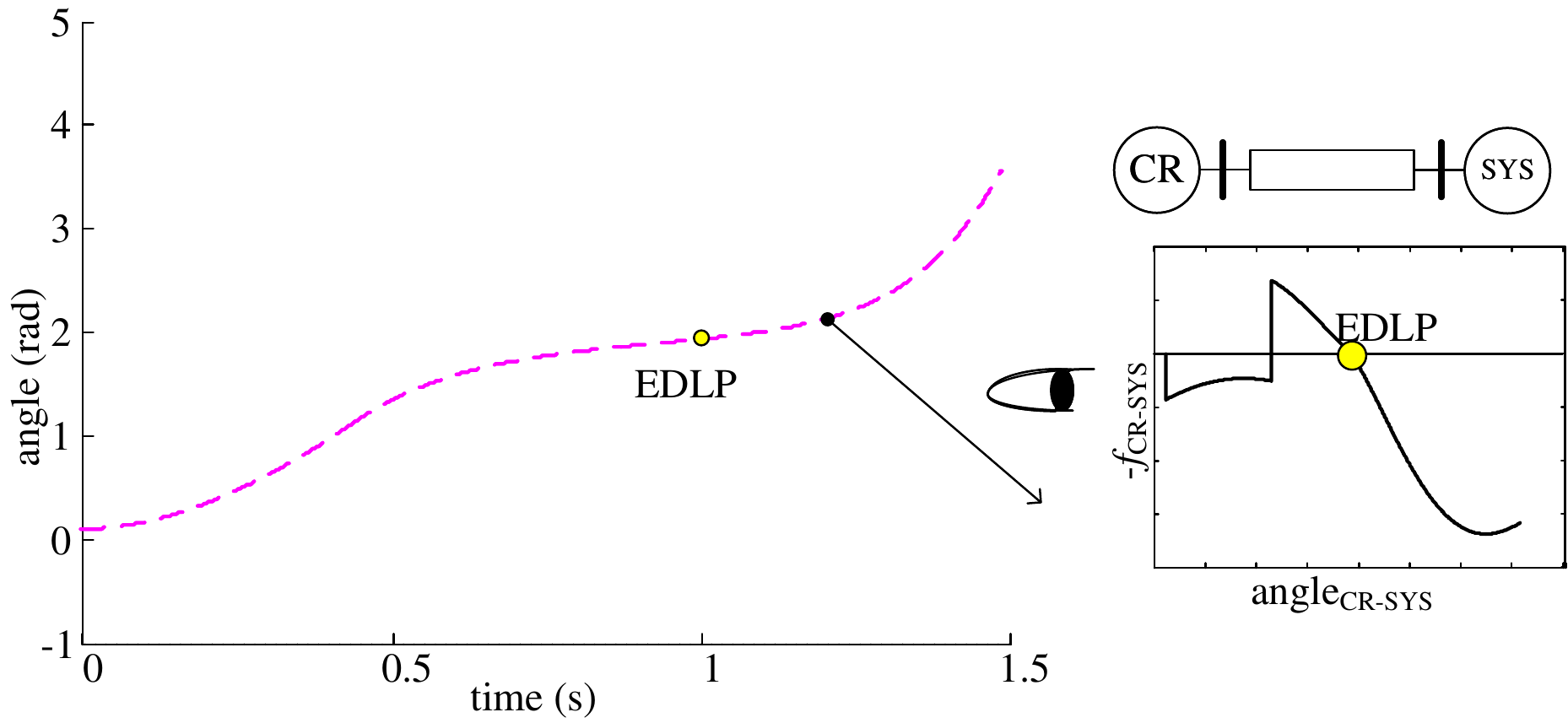}
  \caption{Mechanisms of the real equivalent machine [TS-1, bus-4, 0.447 s].} 
  \label{fig10}  
\end{figure}

 The mechanism of Machine-CR is shown in Fig. \ref{fig10}. From the figure, the strict correlation between each EMTR and EMTE is established through the modeling of the equation of motion of the Machine-CR.
\vspace*{-0.5em}
\noindent 3) ``PSEUDO'' SUPERIMPOSED MACHINE
\vspace*{0.5em}
\par Following the analysis in Section \ref{section_IIB}, the superimposed machine is a typical pseudo machine \cite{3}.
\par The superimposed-machine transient energy (SMTE) is denoted as 
\begin{equation}
  \label{equ15}
  V_{\mathrm{SM}\mbox{-}\mathrm{SYS}}=\sum_{i=1}^{n}V_{i\mbox{-}\mathrm{SYS}}
\end{equation}
\par From Eq. (\ref{equ15}), the superimposed machine is defined based on the ``superimposition'' (addition) of all the IMTEs in the system.
The superimposed machine does not have corresponding equation of motion. It does not have trajectory, mass and force on it. In addition, the NEC of the superimposed machine is pseudo because residual SMKE always exists \cite{3}.
\par
\begin{figure}[H]
  \centering
  \includegraphics[width=0.45\textwidth,center]{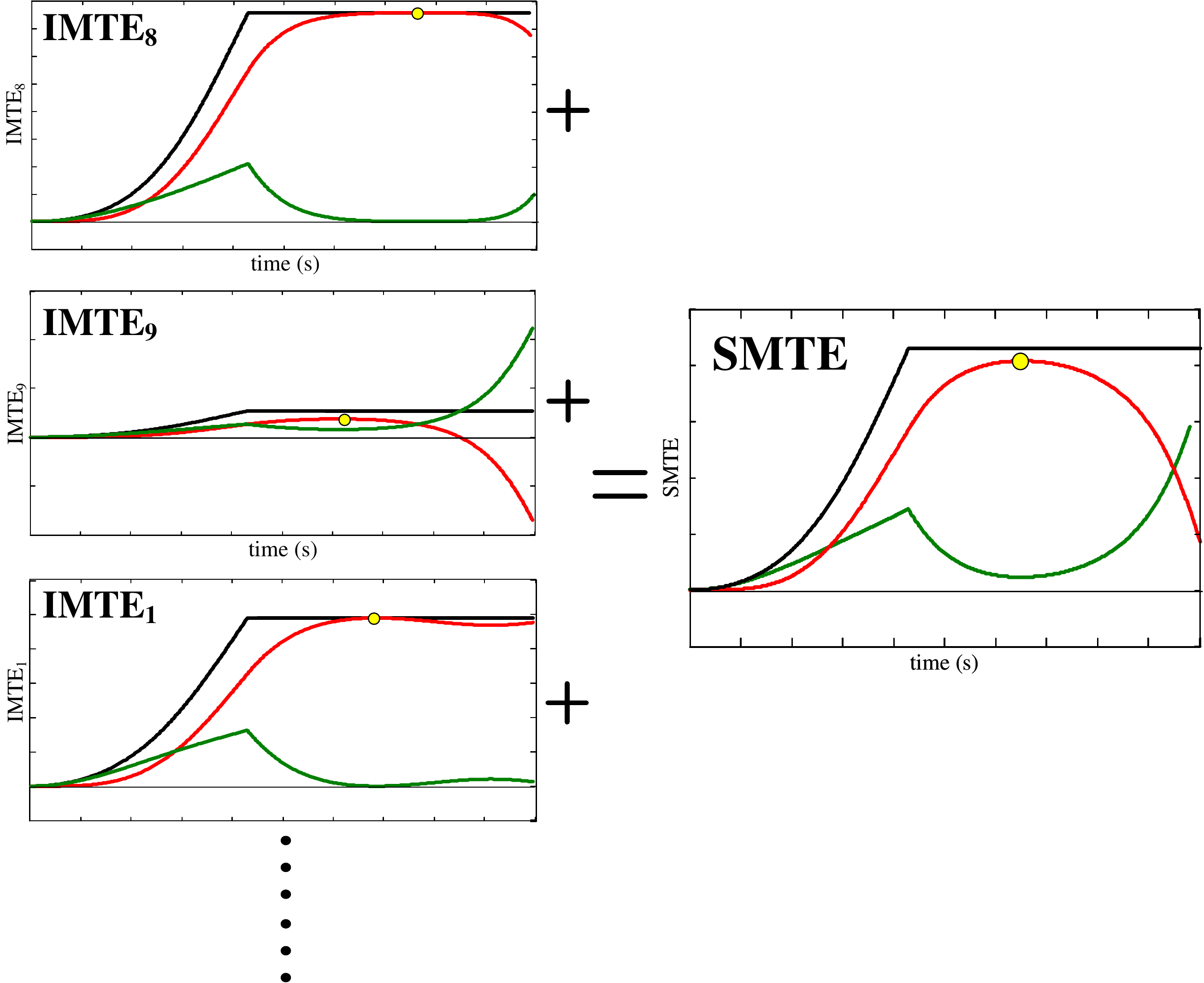}
  \caption{Mechanisms of the superimposed machine [TS-1, bus-2, 0430 s].} 
  \label{fig11}  
\end{figure}

 The mechanism of the superimposed machine is shown in Fig. \ref{fig11}. From the figure, the SMTE is created through the energy superimposition of all machines in the system. The superimposed machine is a pseudo machine without any equation of motion. Therefore, the original system trajectory and the SMTE do not match.
\vspace*{-0.5em}

\subsection{``INDIVIDUAL'' TO “EQUIVALENT” MACHINE TRANSFORAMTION} \label{section_IIIC}
From the analysis in Section \ref{section_IIIB}, following the classifications of machines, both individual machine and equivalent machine are real machines because they both have equation of motions.
The equivalent machine is just depicted in an ``motion equivalence'' manner of the individual machine. Comparatively, the superimposed machine is defined as the ``energy superimposition'' of all individual machines in the system. This energy-computation based superimposed machine is a pseudo machine without any equation of motion.
\par Based on the fact that the pseudo superimposed machine is meaningless to be used in TSA, the machine transformation can only become meaningful from the ``real'' individual machine to the ``real'' equivalent machine. In the following sections, three types of machine transformations, i.e., the ``original definition'' based machine transformation, the ``energy-correction'' based machine transformation and the ``trajectory-correction'' based machine transformation are provided. This may fundamentally clarify the relationship between the individual machine and the equivalent machine. 

\section{MACHINE TRANSFORMATION THROUGH ORIGINAL DEFINITION} \label{section_IV}
\subsection{EXPECTED TRANSFORMATION AFTER MOTION EQUIVALENCE}  \label{section_IVA}
Following the original definition of the equivalent machine, the action of the machine transformation is given as
\vspace*{0.5em}
\\ Action (motion equivalence): The equivalent machine is formed through motion equivalence of all machines inside each group.
\vspace*{0.5em}
\par Based on this motion equivalence, the expected machine transformation is given as below
\vspace*{0.5em}
\\ Trajectory transformation: The original system trajectory becomes the equivalent system trajectory.
\\ Energy transformation: The IMTE becomes the EMTE.
\vspace*{0.5em}
\par
\begin{figure}[H]
  \centering
  \includegraphics[width=0.52\textwidth,center]{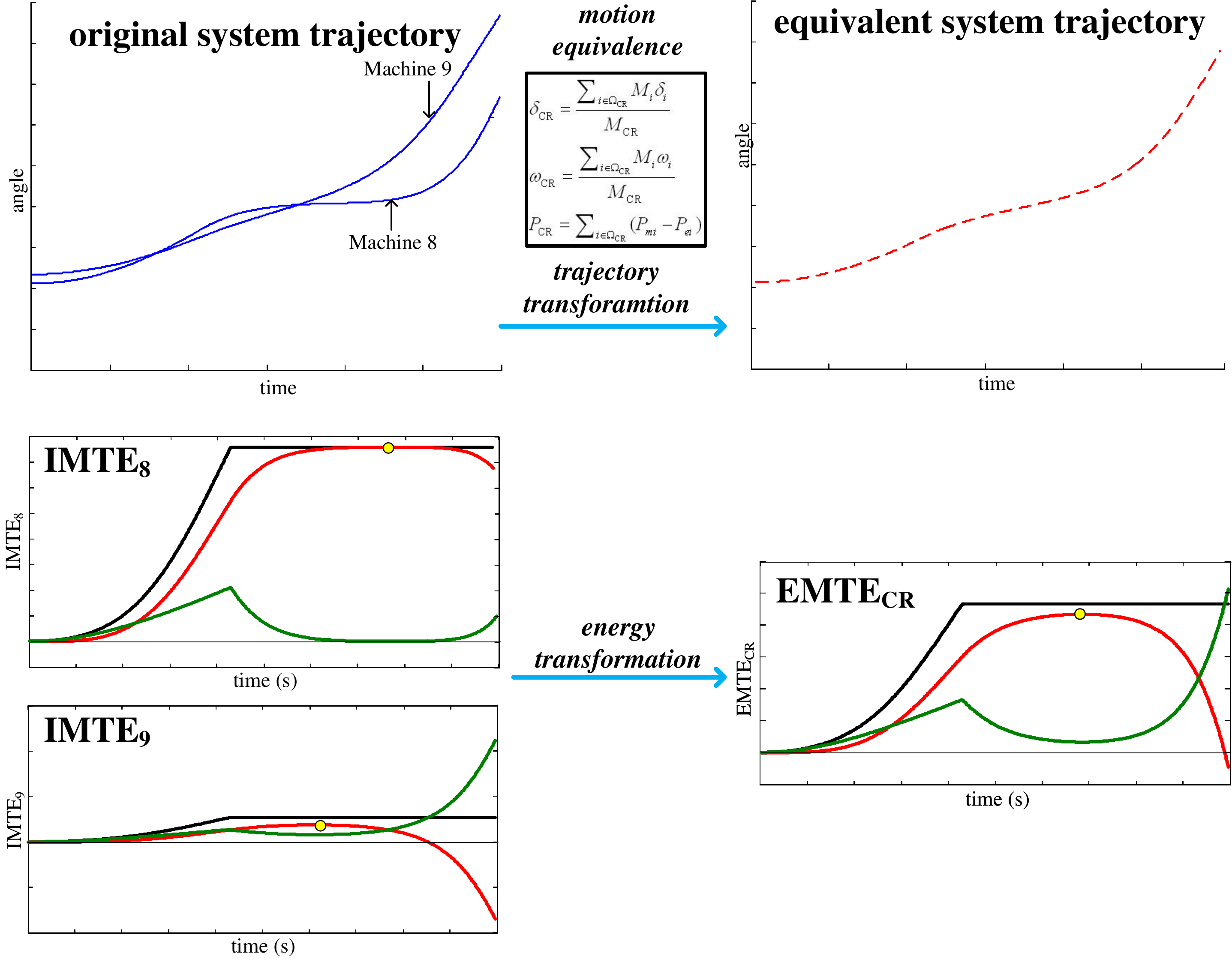}
  \caption{Expected machine transformation after motion equivalence [TS-1, bus-2, 0430 s].} 
  \label{fig12}  
\end{figure}

 The expected machine transformation through motion equivalence is shown in Fig. \ref{fig12}.
The IMTEs of Machines 8 and 9 are shown in Figs. \ref{fig13} (a) and (b), respectively. $\Omega_{\mathrm{CR}}$ is \{Machine 8, Machine 9\}.
The $\text{EMTE}_{\text{CR}}$ is shown in Fig. \ref{fig14}. The original system trajectory and the equivalent system trajectory were already given in Ref. \cite{4}.
\vspace*{-1em}
\begin{figure} [H]
  \centering 
  \subfigure[]{%
  \label{fig13a}
    \includegraphics[width=0.42\textwidth]{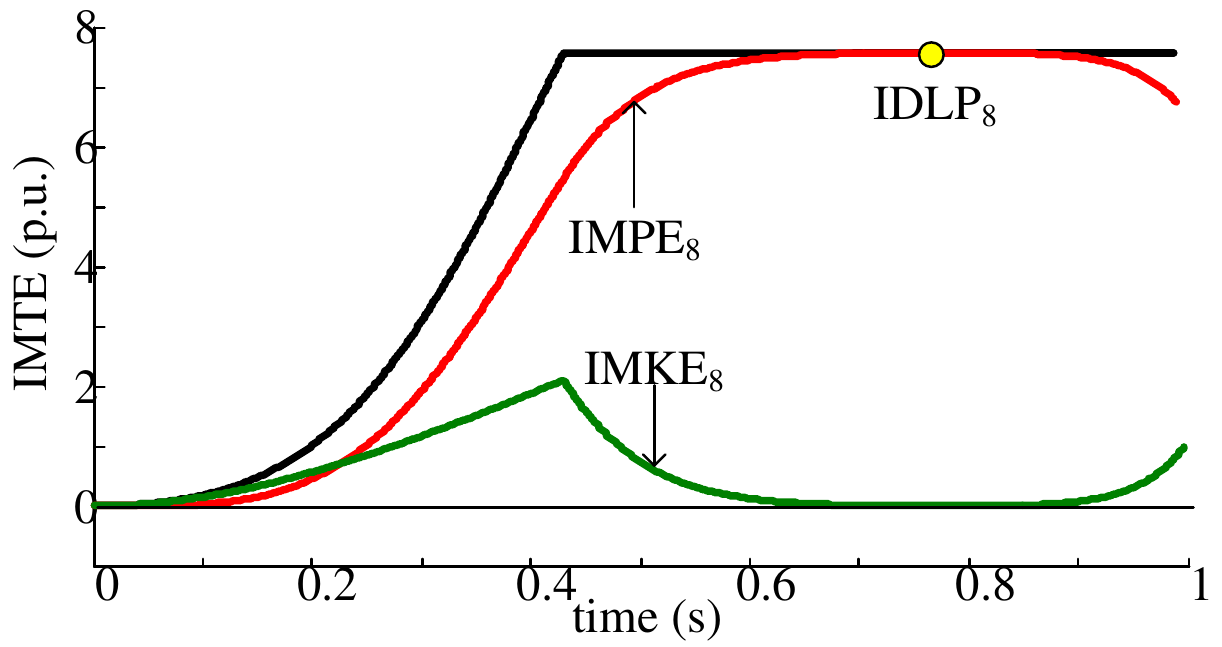}}%
\end{figure} 
\vspace*{-2em}
\addtocounter{figure}{-1}       
\begin{figure} [H]
  \addtocounter{figure}{1}      
  \centering 
  \subfigure[]{%
    \label{fig13b}
    \includegraphics[width=0.42\textwidth]{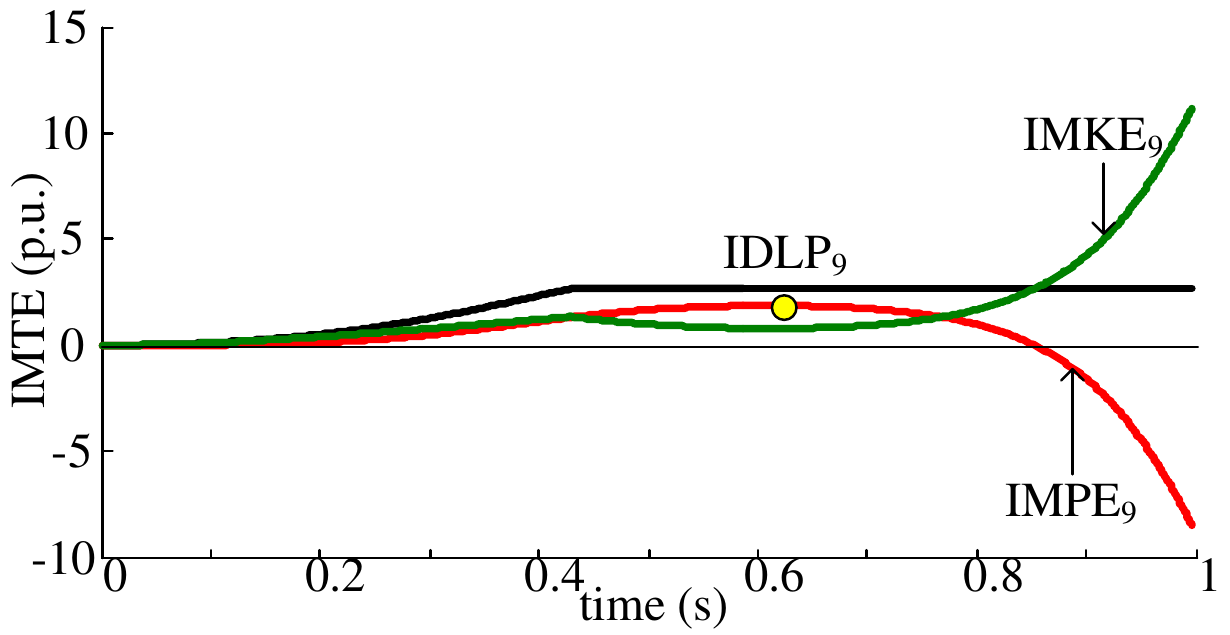}}%
  \caption{Transient energy conversion inside each individual machine [TS-1, bus-2, 0430 s]. (a) Machine 8. (b) Machine 9.}%
  \label{fig13}
\end{figure}
\vspace*{-1em}
\begin{figure}[H]
  \centering
  \includegraphics[width=0.45\textwidth,center]{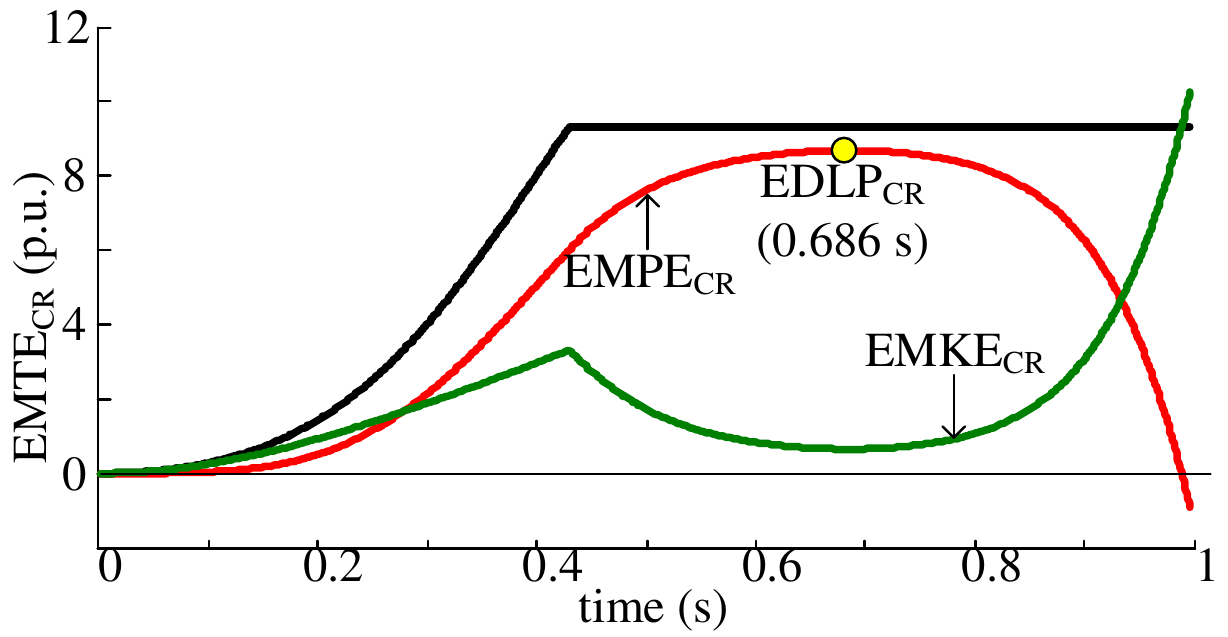}
  \caption{Transient energy conversion inside Machine-CR [TS-1, bus-2, 0430 s].} 
  \label{fig14}  
\end{figure}
\vspace*{-1em}

\subsection{SUCCESS OF MACHINE TRANSFORMATION} \label{section_IVB}
\vspace*{0.5em}
\noindent 1) TRAJECTORY TRANSFORMATION
\vspace*{0.5em}
\par The motion equivalence of Machine-CR is given as
\begin{equation}
  \label{equ16}
  \left\{\begin{array}{l}
    \delta_{\mathrm{CR}\mbox{-}\mathrm{SYS}}=\frac{\sum_{i \in \Omega_{\mathrm{CR}}} M_{i} \delta_{i\mbox{-}\mathrm{SYS}}}{M_{\mathrm{CR}}} \\
    \\
    \omega_{\mathrm{CR}\mbox{-}\mathrm{SYS}}=\frac{\sum_{i \in \Omega_{\mathrm{CR}}} M_{i} \omega_{i\mbox{-}\mathrm{SYS}}}{M_{\mathrm{CR}}} \\
    \\
    f_{\mathrm{CR}\mbox{-}\mathrm{SYS}}=\sum_{i \in \Omega_{\mathrm{CR}}} f_{i\mbox{-}\mathrm{SYS}}
    \end{array}\right.
\end{equation}
\par Following Eq. (\ref{equ16}), $\delta_{\mathrm{CR}\mbox{-}\mathrm{SYS}}$ is established through the equivalence of $\delta_{i\mbox{-}\mathrm{SYS}}$. Therefore, the trajectory transformation is realized through motion equivalence.
\vspace*{0.5em}
\\ 2) ENERGY TRANSFORMATION
\vspace*{0.5em}
\par Based on the motion equivalence as given in Eq. (\ref{equ16}), the equation of motion of Machine-CR is depicted as
\begin{equation}
  \label{equ17}
  \left\{\begin{array}{l}
    \frac{d \delta_{\mathrm{CR}\mbox{-}\mathrm{SYS}}}{d t}=\omega_{\mathrm{CR}\mbox{-}\mathrm{SYS}} \\
    \\
    M_{\mathrm{CR}} \frac{d \omega_{\mathrm{CR}\mbox{-}\mathrm{SYS}}}{d t}=f_{\mathrm{CR}\mbox{-}\mathrm{SYS}}
    \end{array}\right.
\end{equation}
\par From Eq. (\ref{fig17}), EMTE is given in a Newtonian energy manner, as given in Eq. (\ref{equ14}). Therefore, the energy transformation is realized through motion equivalence.
\par From all the analysis above, the motion equivalence based machine transformation is depicted as
\vspace*{0.5em}
\par \textit{Both the trajectory transformation and the energy transformation are realized, because Machine-CR has equivalent equation of motion}.
\vspace*{0.5em}
\par This deduction is also the reflection of the original definition of the equivalent machine.

\subsection{OTHER TWO CONJECTURES ABOUT MACHINE TRANSFORMATION} \label{section_IVC}
\vspace*{0.5em}
\noindent 1) ``REAL'' INNER-GROUP MACHINE
\vspace*{0.5em}
\par Different from original definition of the machine transformation, the system engineer also focuses on the ``removal'' (subtraction) of the effects of the inner-group machines from the original system.
\par
\begin{figure}[H]
  \centering
  \includegraphics[width=0.42\textwidth,center]{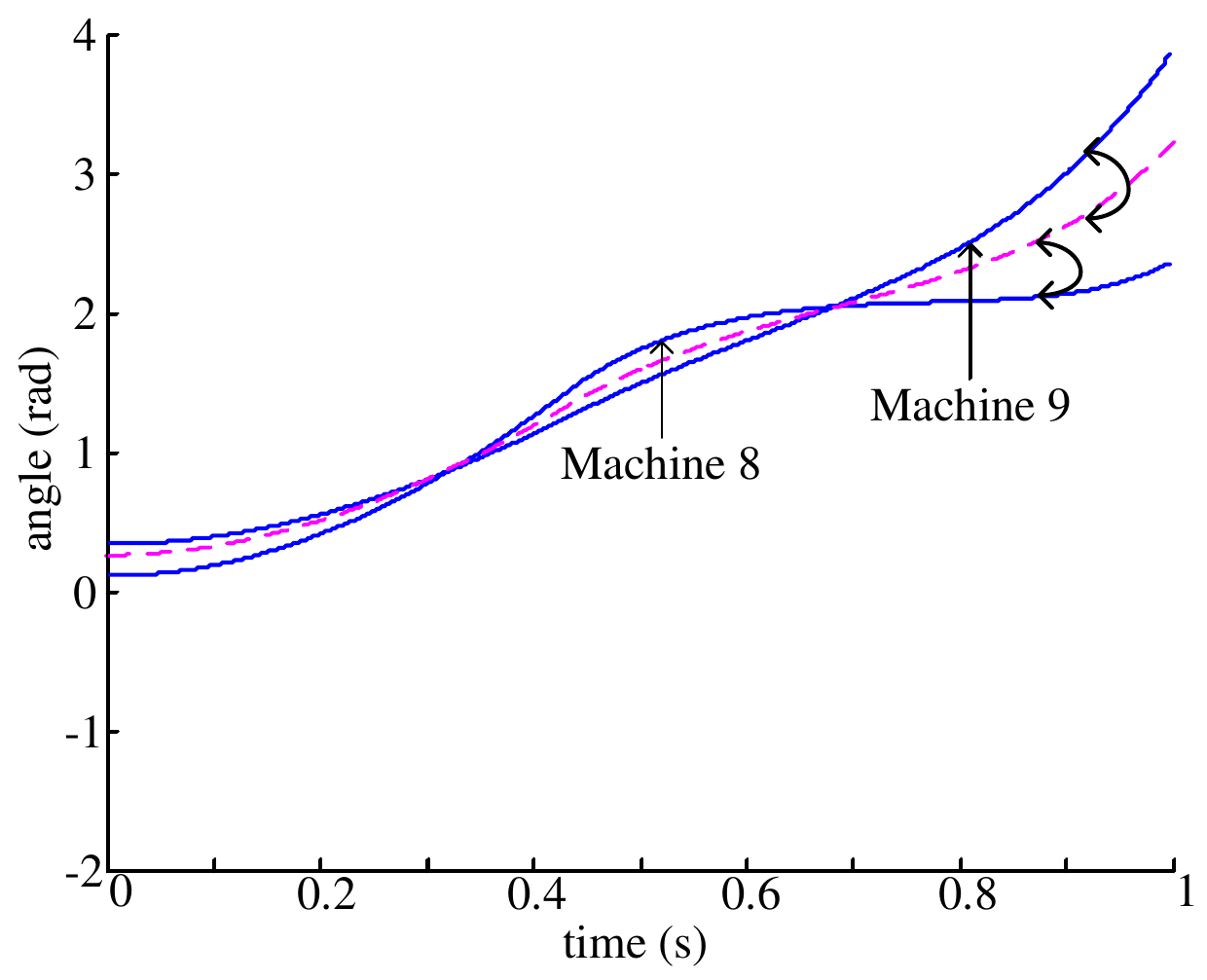}
  \caption{Inner-group machine motions inside Group-CR [TS-1, bus-2, 0.430 s].} 
  \label{fig15}  
\end{figure}

 Following the analysis in Ref. \cite{5}, the inner-group machine is the difference between the original system and the equivalent system. The mechanisms of the inner-group machine are shown in Fig. \ref{fig15}.
\vspace*{-0.5em}
The equation of motion of each inner-group machine is depicted as \cite{5}
\begin{equation}
  \label{equ18}
  \left\{\begin{array}{l}
    \frac{d \delta_{i\mbox{-}\mathrm{CR}}}{d t}=\omega_{i\mbox{-}\mathrm{CR}} \\
    \\
    M_{i} \frac{d \omega_{i\mbox{-}\mathrm{CR}}}{d t}=f_{i\mbox{-}\mathrm{CR}}
    \end{array}\right.
\end{equation}
where
\vspace*{0.5em}
\\
$\left\{\begin{array}{l}
  \omega_{i\mbox{-}\mathrm{CR}}=\omega_{i\mbox{-}\mathrm{SYS}}-\omega_{\mathrm{CR}\mbox{-}\mathrm{SYS}} \\
  \\
  f_{i\mbox{-}\mathrm{CR}}=f_{i\mbox{-}\mathrm{SYS}}-\frac{M_{i}}{M_{\mathrm{CR}}} f_{\mathrm{CR}\mbox{-}\mathrm{SYS}}
  \end{array}\right.$
\vspace*{0.5em}
\par Based on Eq. (\ref{fig18}), the inner-group machine transient energy (IGMTE) is defined in a typical Newtonian energy manner. The IGMTE is defined as
\begin{equation}
  \label{equ19}
  V_{i\mbox{-}\mathrm{CR}}=V_{K E i\mbox{-}\mathrm{CR}}+V_{P E i\mbox{-}\mathrm{CR}}
\end{equation}
where
\vspace*{0.5em}
\\
$\left\{\begin{array}{l}
  V_{K E i\mbox{-}\mathrm{CR}}=\frac{1}{2} M_{i} \omega_{i\mbox{-}\mathrm{CR}}^{2} \\
  \\
  V_{P E i\mbox{-}\mathrm{CR}}=\int_{\delta_{i\mbox{-}\mathrm{CR}}^{s}}^{\delta_{i\mbox{-}\mathrm{CR}}}\left[-f_{i\mbox{-}\mathrm{CR}}^{(P F)}\right] d \delta_{i\mbox{-}\mathrm{CR}}
  \end{array}\right.$
\vspace*{0.5em}
\par From Eqs. (\ref{equ18}) and (\ref{equ19}), following the machine classifications as analyzed in Section \ref{section_III}, the inner-group machine has its equation of motion. Therefore, the inner-group machine is also a ``real'' machine that is defined in the COI-CR reference. 
\par The IGMTEs of Machines 8 and 9 are shown in Figs. \ref{fig16} (a) and (b), respectively.
\begin{figure} [H]
  \centering 
  \subfigure[]{%
  \label{fig16a}
    \includegraphics[width=0.46\textwidth]{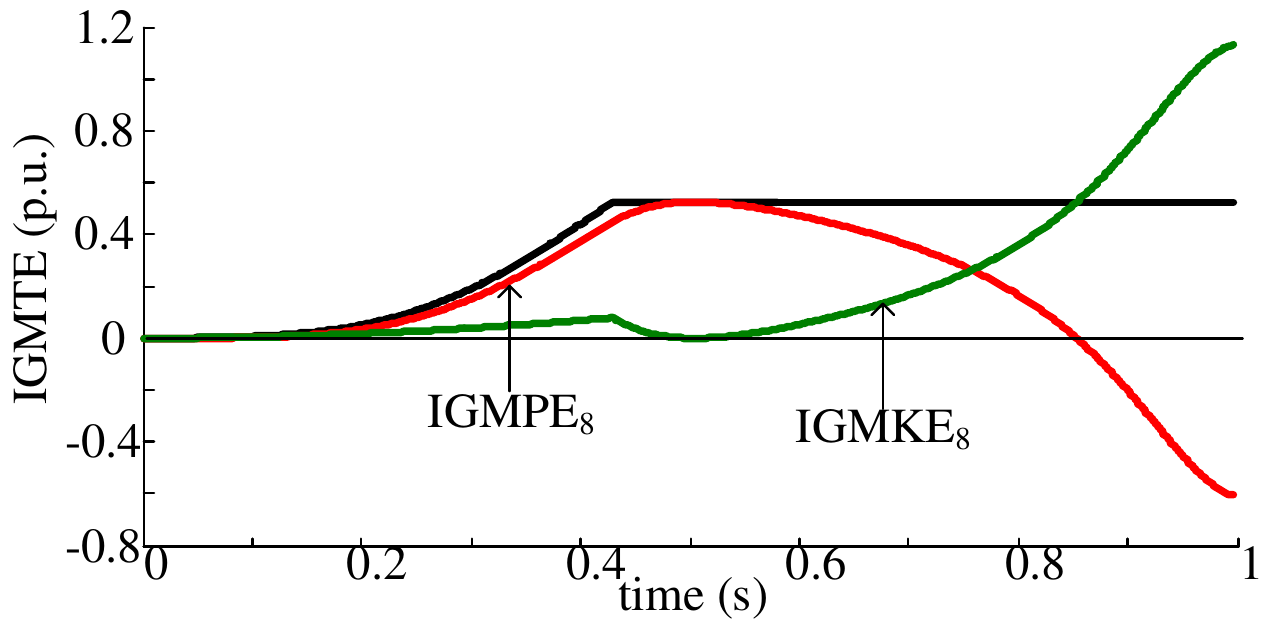}}%
\end{figure} 
\vspace*{-2em}
\addtocounter{figure}{-1}       
\begin{figure} [H]
  \addtocounter{figure}{1}      
  \centering 
  \subfigure[]{%
    \label{fig16b}
    \includegraphics[width=0.46\textwidth]{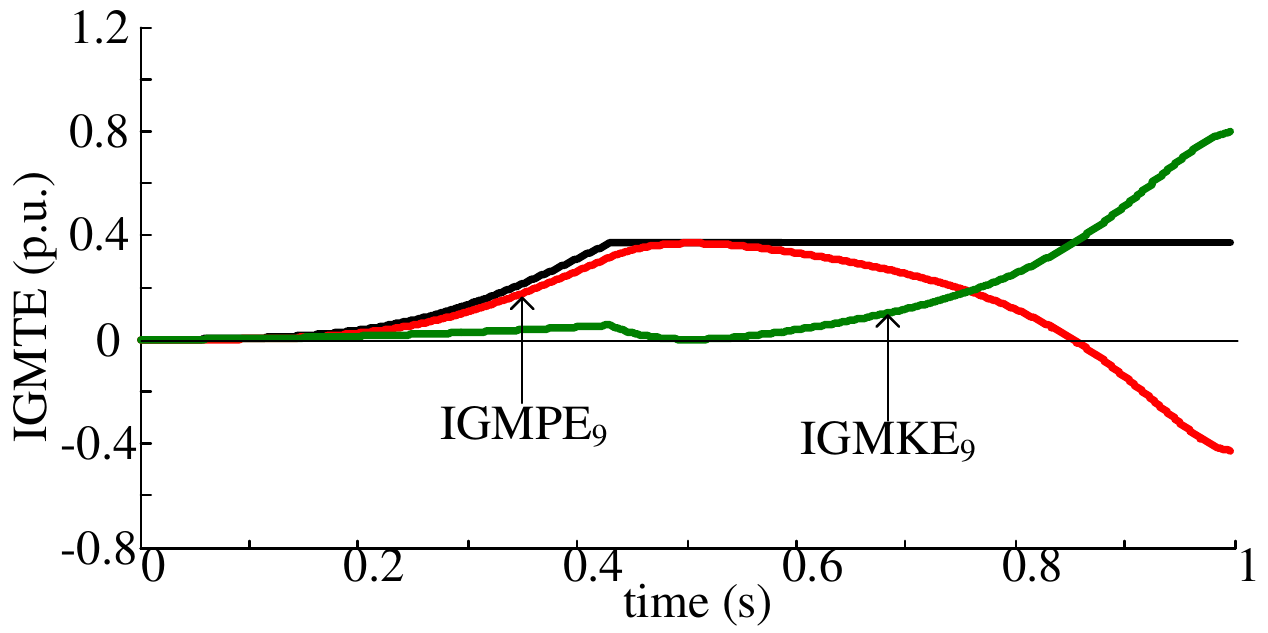}}%
  \caption{IGMTE [TS-1, bus-2, 0.430 s]. (a) $\text{IGMTE}_8$. (b) $\text{IGMTE}_9$.}%
  \label{fig16}
\end{figure}
\vspace*{-0.5em}
\noindent 2) ENERGY CORRECTION AND TRAJECTORY CORRECTION
\vspace*{0.5em}
\par From Figs. \ref{fig15} and \ref{fig16}, the IGMTR and IGMTE represent the ``trajectory'' and ``energy'' of the inner-group machine, respectively. Further, since the inner-group machine motions are the differences between the original system and the equivalent machine, two conjectures about machine transformations are naturally formed. The two conjectures are given as below
\vspace*{0.5em}
\\ CONJECTURE-I: The IGMTE should be ``corrected'' (subtracted) from the IMTE. Then, machine transformation will be realized (energy correction perspective).
\\ CONJECTURE-II: The IGMTR should be subtracted from the IMTR. Then, machine transformation will be realized (trajectory correction perspective).
\vspace*{0.5em}
\par The two conjectures also form two different machine transformations, i.e., machine transformation through energy correction and the machine transformation through trajectory correction. Detailed analysis is given in the following sections.

\section{MACHINE TRANSFORMATION THROUGH ENERGY CORRECTIONS} \label{section_V}
\subsection{EXPECTED STATE AFTER ENERGY CORRECTIONS}  \label{section_VA}
Following CONJECTURE-I, the action of this machine transformation is given as
\vspace*{0.5em}
\\ Action (energy correction): The IGMTE is corrected from each IMTE inside Group-CR.
\vspace*{0.5em}
\par Against this background, the energy-correction based individual machine (ECIM) is obtained (the analysis is given in Section \ref{section_VB}). The expected machine transformation (including both trajectory transformation and energy transformation) is given as below
\vspace*{0.5em}
\\ Energy transformation: the energy transformation is assumed to be realized after the superimposition of all the ECIM transient energy (ECIMTE) in the system.
\\ Trajectory transformation: The trajectory of ECIM becomes the EMTR.
\vspace*{0.5em}
\par
\begin{figure}[H]
  \centering
  \includegraphics[width=0.5\textwidth,center]{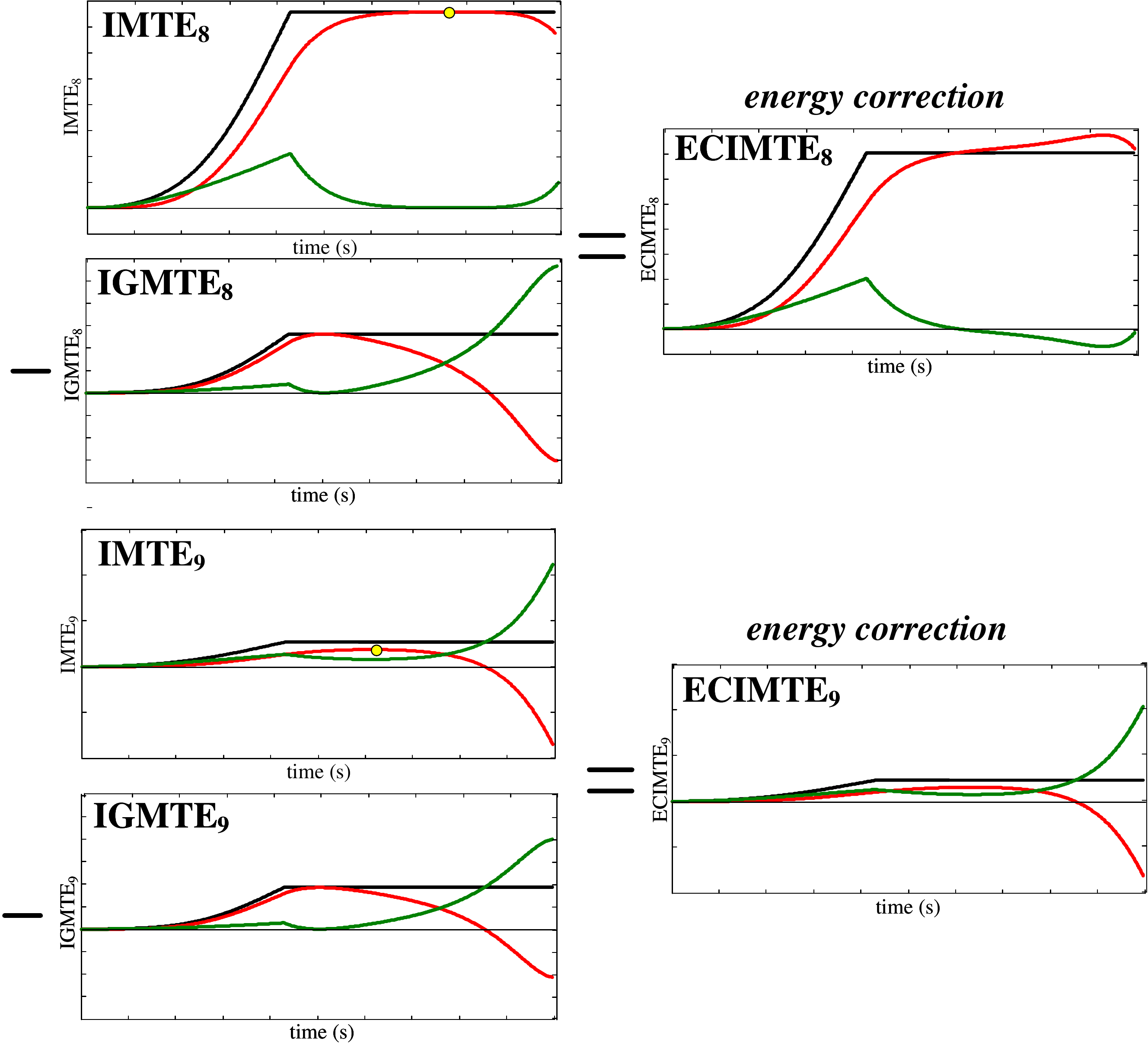}
  \caption{Formation of ECIM [TS-1, bus-2, 0430 s].} 
  \label{fig17}  
\end{figure}

 The formation of ECIM is shown in Fig. \ref{fig17}. The expected system state after energy corrections are shown in Figs. \ref{fig18} (a) and (b), respectively.
\vspace*{-1em}
\begin{figure} [H]
  \centering 
  \subfigure[]{%
  \label{fig18a}
    \includegraphics[width=0.45\textwidth]{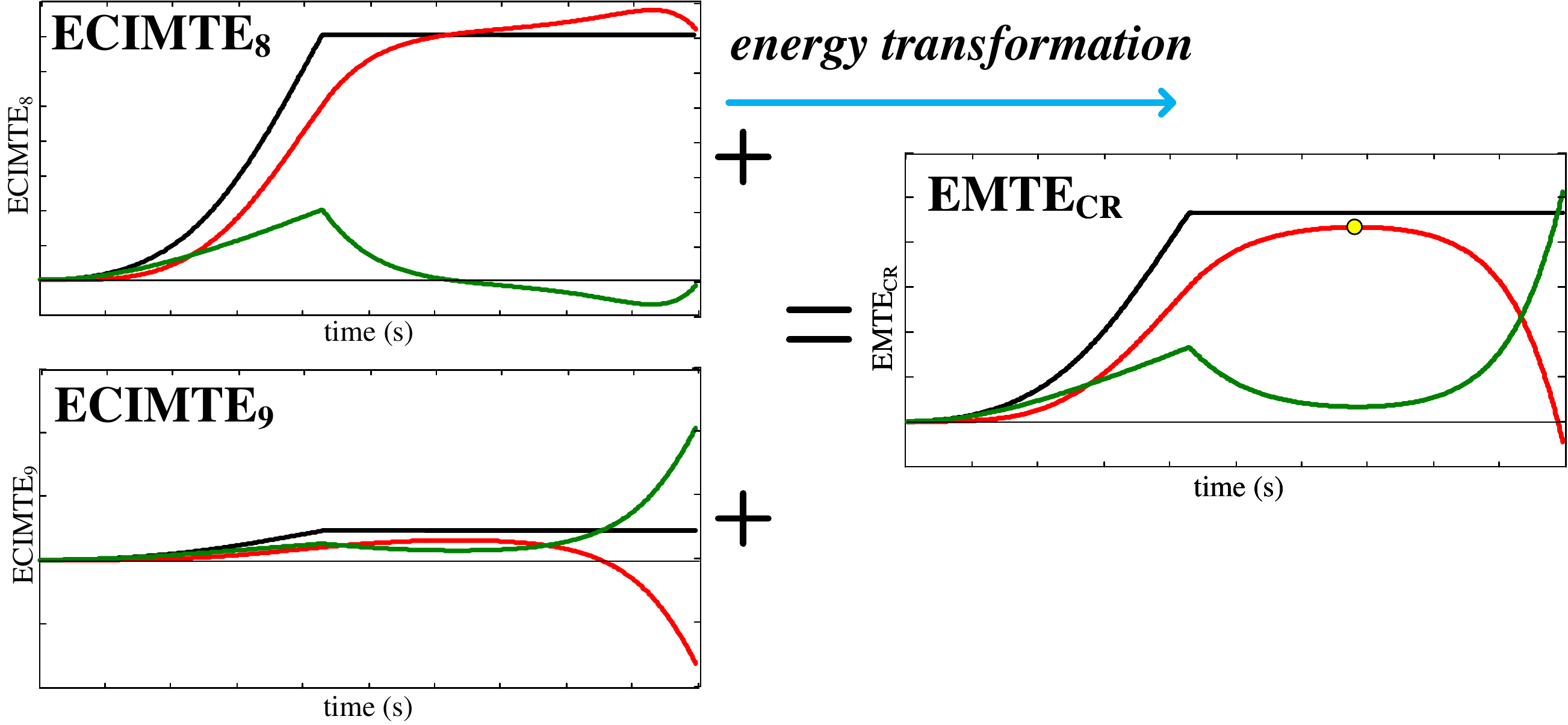}}%
\end{figure} 
\vspace*{-2em}
\addtocounter{figure}{-1}       
\begin{figure} [H]
  \addtocounter{figure}{1}      
  \centering 
  \subfigure[]{%
    \label{fig18b}
    \includegraphics[width=0.45\textwidth]{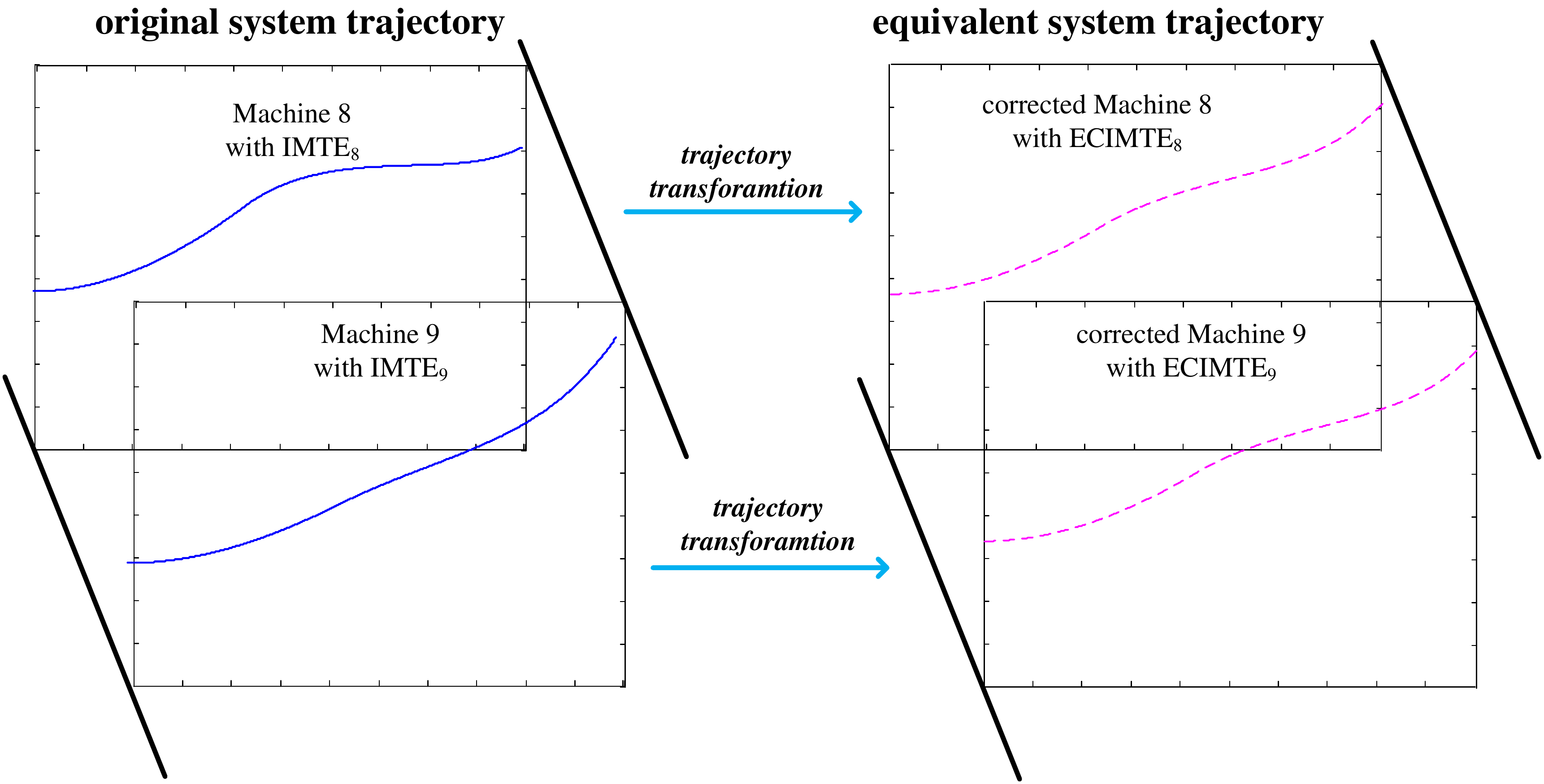}}%
  \caption{Expected system state after energy corrections. (a) Energy transformation. (b) Trajectory transformation.}%
  \label{fig18}
\end{figure}
\vspace*{-0.5em}
From Figs. \ref{fig17} and \ref{fig18}, in brief, the system engineer first focuses on the modeling of the ECIM. Then, the energy transformation will be realized through the superimposition of all ECIMTEs. Meanwhile, Then, the trajectory transformation is also realized in each ECIM.

\subsection{``PSEUDO'' ECIM} \label{section_VB}
Following the IMTE as in Eq. (\ref{equ12}) and IGMTE as in Eq. (\ref{equ19}), the ECIMTE is given as
\begin{equation}
  \label{equ20}
  V_{i\mbox{-}\mathrm{SYS}}^{(\mathrm{EC})}=V_{K E i\mbox{-}\mathrm{SYS}}^{(\mathrm{EC})}+V_{P E i\mbox{-}\mathrm{SYS}}^{(\mathrm{EC})}
\end{equation}
where
\vspace*{0.5em}
\\$\left\{\begin{array}{l}
  V_{K E i\mbox{-}\mathrm{SYS}}^{(\mathrm{EC})}=V_{K E i\mbox{-}\mathrm{SYS}}-V_{K E i\mbox{-}\mathrm{CR}} \\
  \\
  V_{P E i\mbox{-}\mathrm{SYS}}^{(\mathrm{EC})}=V_{P E i\mbox{-}\mathrm{SYS}}-V_{P E i\mbox{-}\mathrm{CR}}
  \end{array}\right.$
\vspace*{0.5em}
\par Also based on Eqs. (\ref{equ12}) and (\ref{equ19}), the following holds
\begin{equation}
  \label{equ21}
  V_{i\mbox{-}\mathrm{SYS}}^{(\mathrm{ECIM})}=V_{i\mbox{-}\mathrm{SYS}}-V_{i\mbox{-}\mathrm{CR}}
\end{equation}
\par In Eq. (\ref{equ21}), the ECIM is obtained through the subtraction of IGMTE from IMTE. Following the analysis in Section \ref{section_III}, ECIM is a typical pseudo machine without equation of motion.
\par The characteristics of the energy conversion inside the ECIM are given as below
\vspace*{0.5em}
\\(i) ECIMTE remains conservative, because both the IMTE and IGMTE remain conservative, as in Eq. (\ref{equ20}).
\\ (ii) ECIMTE is lower than IMTE because IGMTE is positive.
\\ (iii) the transient energy conversion inside each pseudo ECIM cannot satisfy NEC characteristic.
\vspace*{0.5em}
\par For the case in Fig. \ref{fig15}, the ECIMTE of each ECIM are shown in Figs. \ref{equ19} (a) and (b), respectively.
The IMTE and IGMTE of the two machines were already given in Figs. \ref{fig13} and \ref{fig16}.
\begin{figure} [H]
  \centering 
  \subfigure[]{%
  \label{fig19a}
    \includegraphics[width=0.45\textwidth]{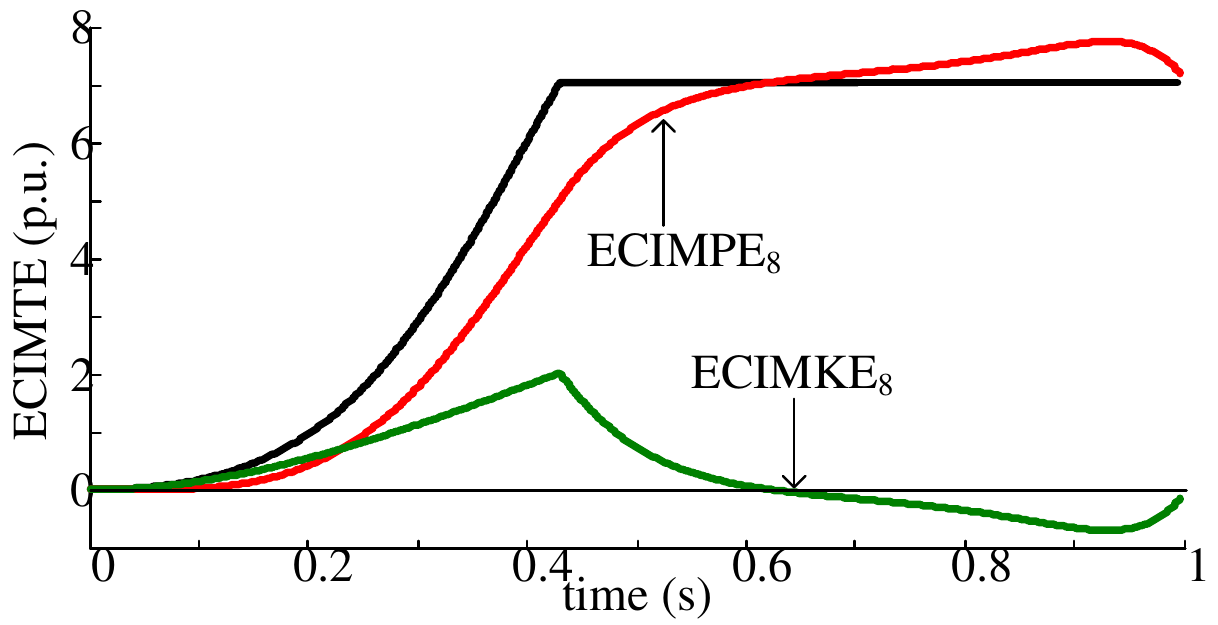}}%
\end{figure} 
\vspace*{-2em}
\addtocounter{figure}{-1}       
\begin{figure} [H]
  \addtocounter{figure}{1}      
  \centering 
  \subfigure[]{%
    \label{fig19b}
    \includegraphics[width=0.45\textwidth]{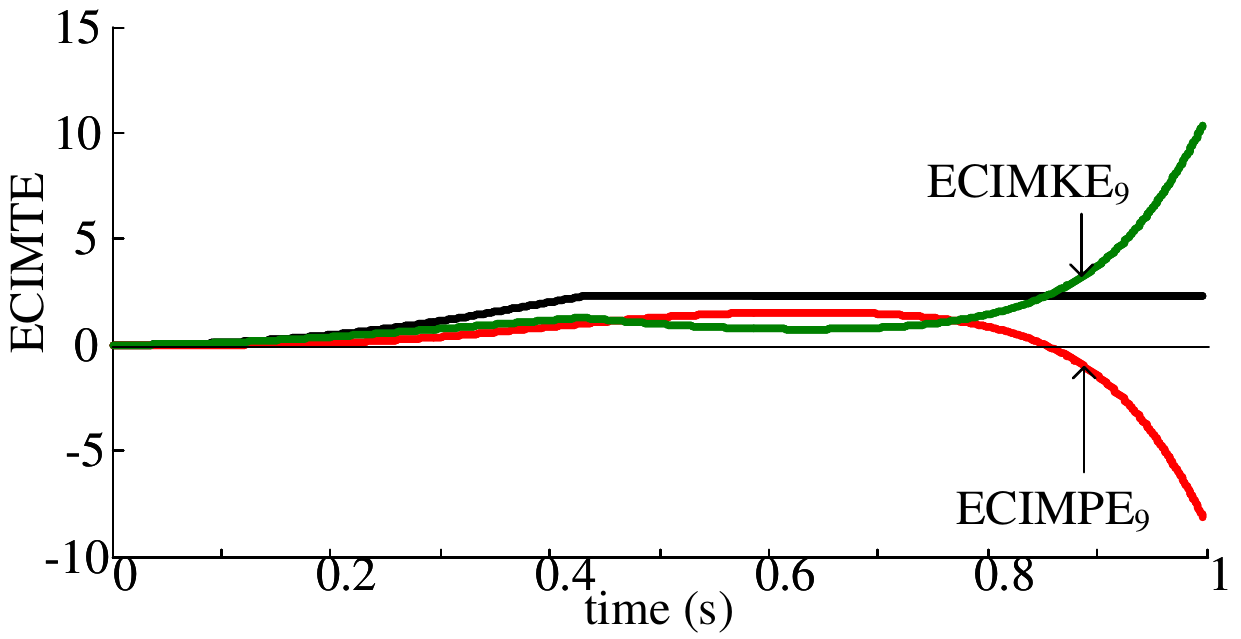}}%
  \caption{CIMTE after transient energy corrections [TS-1, bus-2, 0.430 s]. (a) $\text{CIMTE}_8$. (b) $\text{CIMTE}_9$.}%
  \label{fig19}
\end{figure}
\vspace*{-1.5em}
\subsection{``MATHEMATICAL'' SUCCESS OF ENERGY TRANSFORMATION} \label{section_VC}
Following the ECIMTE of each pseudo ECIM, the superimposition of all the ECIMTEs inside Group-CR is defined as
\begin{equation}
  \label{equ22}
  V_{\mathrm{SM}\mbox{-}\mathrm{SYS}}^{(\mathrm{EC})}=V_{K E \mathrm{SM}\mbox{-}\mathrm{SYS}}^{(\mathrm{EC})}+V_{P E \mathrm{SM}\mbox{-}\mathrm{SYS}}^{(\mathrm{EC})}=\sum_{k=1}^{n} V_{i\mbox{-}\mathrm{SYS}}^{(\mathrm{EC})}
\end{equation}
where
\vspace*{0.5em}
\\$\left\{\begin{array}{l}
  V_{K E \mathrm{SM}\mbox{-}\mathrm{SYS}}^{(\mathrm{EC})}=\sum V_{K E i\mbox{-}\mathrm{SYS}}^{(\mathrm{EC})} \\
  \\
  V_{PE \mathrm{SM}\mbox{-}\mathrm{SYS}}^{(\mathrm{EC})}=\sum V_{P E i\mbox{-}\mathrm{SYS}}^{(\mathrm{EC})}
  \end{array}\right.$
\vspace*{0.5em}
\par Further, the following equation can be obtained
\begin{equation}
  \label{equ23}
  V_{\mathrm{SM}\mbox{-}\mathrm{SYS}}^{(\mathrm{EC})}=V_{\mathrm{CR}\mbox{-}\mathrm{SYS}}
\end{equation}
\par Eq. (\ref{equ23}) indicates that the EMTE of Machine-CR is ``mathematically'' obtained through the superimposition of all the ECIMTEs inside Group-CR.
\par The proof about the mathematical equivalence in Eq. (\ref{equ23}) is quite complicated. Detailed proof is given as below
\vspace*{0.2em}\\
\textit{Step-1}: $V_{KE\mathrm{SM}\mbox{-}\mathrm{SYS}}^{(\mathrm{EC})}$ can be depicted as
\vspace*{0.2em}
\begin{equation}
  \begin{split}
    V_{K E\mathrm{SM}\mbox{-}\mathrm{SYS}}^{(\mathrm{EC})}&=\sum V_{K Ei\mbox{-}\mathrm{SYS}}^{(\mathrm{EC})}\\
    &=\sum\left(V_{K Ei\mbox{-}\mathrm{SYS}}-V_{K E i\mbox{-}\mathrm{CR}}\right)
  \end{split}
  \label{equ24}
\end{equation}
\par In Eq. (\ref{equ24}), the following holds
\begin{equation}
  \label{equ25}
  \left\{\begin{array}{l}
    V_{K E i\mbox{-}\mathrm{CR}}=\frac{1}{2} M_{i}\left(\omega_{i\mbox{-}\mathrm{SYS}}-\omega_{\mathrm{CR}\mbox{-}\mathrm{SYS}}\right)^{2} \\
    \\
    \sum M_{i} \omega_{i\mbox{-}\mathrm{SYS}}=M_{\mathrm{CR}} \omega_{\mathrm{CR}\mbox{-}\mathrm{SYS}} \\
    \\
    \sum M_{i}=M_{\mathrm{CR}}
    \end{array}\right.
\end{equation}
\par Substituting Eq. (\ref{equ25}) into Eq. (\ref{equ24}), one can obtain
\begin{equation}
  \label{equ26}
  \begin{split}
      &V_{KE\mathrm{SM}\mbox{-}\mathrm{SYS}}^{(\mathrm{EC})} =\sum M_{i} \omega_{i\mbox{-}\mathrm{SYS}} \omega_{\mathrm{CR}\mbox{-}\mathrm{SYS}}-\\
      &\sum \frac{1}{2} M_{i} \omega_{\mathrm{CR}\mbox{-}\mathrm{SYS}}^{2}=V_{K E \mathrm{C R}\mbox{-}\mathrm{SYS}}
  \end{split}
\end{equation}
\par Eq. (\ref{equ26}) indicates that the mathematical equivalence of the kinetic energy holds.
\vspace*{0.2em}
\\\textit{Step-2}: $V_{PE\mathrm{SM}\mbox{-}\mathrm{SYS}}^{(\mathrm{EC})}$  can be depicted as
\vspace*{0.2em}
\begin{equation}
  \label{equ27}
  \begin{split}
    V_{PE\mathrm{SM}\mbox{-}\mathrm{SYS}}^{(\mathrm{EC})}&=\sum V_{P E i\mbox{-}\mathrm{SYS}}^{(\mathrm{EC})}\\
    &=\sum\left(V_{P E i\mbox{-}\mathrm{SYS}}-V_{P E i\mbox{-}\mathrm{CR}}\right)
  \end{split}
\end{equation}
\par $\sum V_{PEi\mbox{-}\mathrm{CR}}$ in Eq. (\ref{equ27}) can be depicted as
\begin{equation}
  \label{equ28}
  \begin{split}
    \sum V_{PEi\mbox{-}\mathrm{CR}}&=\sum \int_{\delta_{i\mbox{-}\mathrm{CR}}^{s}}^{\delta_{i\mbox{-}\mathrm{CR}}}\left[-f_{i\mbox{-}\mathrm{SYS}}^{(P F)}+\frac{M_{i}}{M_{\mathrm{CR}}} f_{\mathrm{CR}\mbox{-}\mathrm{SYS}}^{(P F)}\right] d \delta_{i\mbox{-}\mathrm{CR}}\\
    &=\sum \int_{\delta_{i\mbox{-}\mathrm{CR}}^{s}}^{\delta_{i\mbox{-}\mathrm{CR}}}\left[-f_{i\mbox{-}\mathrm{SYS}}^{(P F)}\right] d \delta_{i\mbox{-}\mathrm{CR}}+ \\
    &\sum \int_{\delta_{i\mbox{-}\mathrm{CR}}^{s}}^{\delta_{i\mbox{-}\mathrm{CR}}}\left[\frac{M_{i}}{M_{\mathrm{CR}}} f_{\mathrm{CR}\mbox{-}\mathrm{SYS}}^{(P F)}\right] d \delta_{i\mbox{-}\mathrm{CR}}
  \end{split}
\end{equation}
\par Inside Group-CR, the following holds
\begin{equation}
  \label{equ29}
  \sum M_{i} \delta_{i\mbox{-}\mathrm{CR}}=0 \Rightarrow \sum d M_{i} \delta_{i\mbox{-}\mathrm{CR}}=0
\end{equation}
\par Substituting Eq. (\ref{equ29}) into Eq. (\ref{equ28}) yields 
\begin{equation}
  \label{equ30}
  \sum \int_{\delta_{i\mbox{-}\mathrm{CR}}^{s}}^{\delta_{i\mbox{-}\mathrm{CR}}}\left[\frac{f_{\mathrm{CR}\mbox{-}\mathrm{SYS}}}{M_{\mathrm{CR}}}\right] d M_{i} \delta_{i\mbox{-}\mathrm{CR}}=0
\end{equation}
\par Following Eq. (\ref{equ30}), Eq. (\ref{equ28}) can be re-expressed as
\begin{equation}
  \label{equ31}
  \sum V_{P E i\mbox{-}\mathrm{CR}}=\sum \int_{\delta_{i\mbox{-}\mathrm{CR}}^s}^{\delta_{i\mbox{-}\mathrm{CR}}}\left[-f_{i\mbox{-}\mathrm{SYS}}\right] d \delta_{i\mbox{-}\mathrm{CR}}
\end{equation}
\par Inside Group-CR, the following also holds
\begin{equation}
  \label{equ32}
  \left\{\begin{array}{l}
    \delta_{i\mbox{-}\mathrm{CR}}=\delta_{i\mbox{-}\mathrm{SYS}}-\delta_{\mathrm{CR}\mbox{-}\mathrm{SYS}}\\
    \\
    \sum f_{i\mbox{-}\mathrm{SYS}}=f_{\mathrm{CR}\mbox{-}\mathrm{SYS}}
    \end{array}\right.
\end{equation}
\par Following Eq. (\ref{equ31}), Eq. (\ref{equ32}) can be further denoted as
\begin{equation}
  \label{equ33}
  \small
  \begin{split}
  &\sum V_{P E i\mbox{-}\mathrm{CR}}\\
  =&\sum \int_{\delta_{i\mbox{-}\mathrm{SYS}}^{s}-\delta_{\mathrm{CR}\mbox{-}\mathrm{SYS}}^{s}}^{\delta_{i\mbox{-}\mathrm{SYS}}-\delta_{\mathrm{CR}\mbox{-}\mathrm{SYS}}}\left[-f_{i\mbox{-}\mathrm{SYS}}\right]d\left(\delta_{i\mbox{-}\mathrm{SYS}}-\delta_{\mathrm{CR}\mbox{-}\mathrm{SYS}}\right)\\
  =&\sum \int_{\delta_{i\mbox{-}\mathrm{CYS}}^{s}}^{\delta_{i\mbox{-}\mathrm{SYS}}}\left[-f_{i\mbox{-}\mathrm{SYS}}\right] d \delta_{i\mbox{-}\mathrm{SYS}}-\\
  &\sum \int_{\delta_{\mathrm{CR}-\mathrm{SYS}}^{s}}^{\delta_{\mathrm{CR}\mbox{-}\mathrm{SYS}}}\left[-f_{i\mbox{-}\mathrm{SYS}}\right] d \delta_{\mathrm{CR}\mbox{-}\mathrm{SYS}}\\
  =&\sum V_{P E i\mbox{-}\mathrm{SYS}}-V_{P E \mathrm{C R}\mbox{-}\mathrm{SYS}}
  \end{split}
\end{equation}
\par Eq. (\ref{equ33}) indicates that the mathematical equivalence of the kinetic energy holds.
\par The analysis above is the systematic proof of the well-known conjecture of the transient energy correction in the history of equivalent machine \cite{8}, \cite{9}.
\par
\begin{figure}[H]
  \centering
  \includegraphics[width=0.42\textwidth,center]{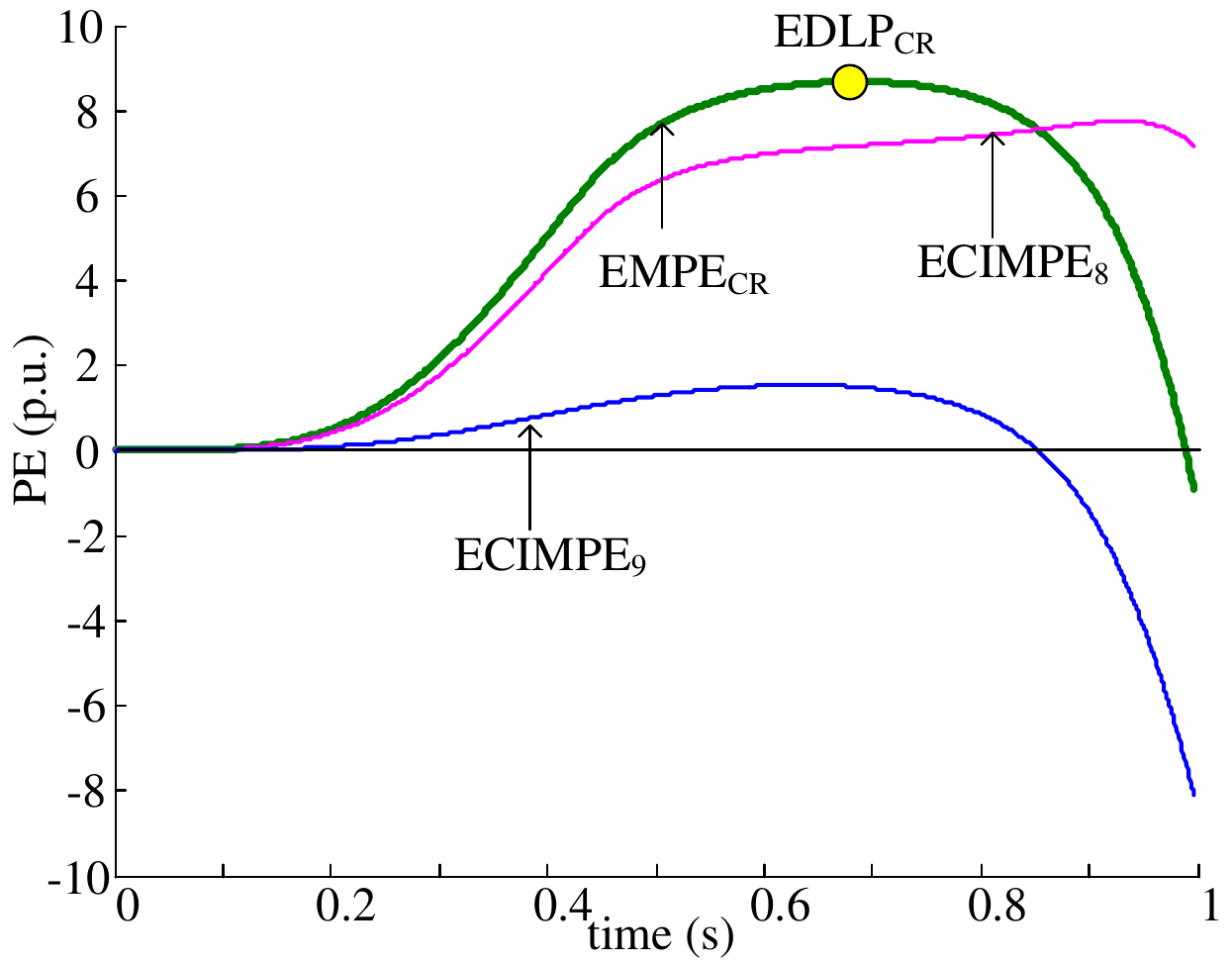}
  \caption{EMPE after the superimposition of the ECIMPEs [TS-1, bus-2, 0.430 s].} 
  \label{fig20}  
\end{figure}

 A tutorial example is given as below. The EMPE of Machine-CR after the superimposition of all the ECIMPEs inside Group-CR is shown in Fig. \ref{fig20}.
\vspace*{-0.5em}
At this stage, Machine-CR is mathematically obtained through the superimposition of the ECIMTEs, as in Fig. \ref{fig19}. Then, it ``seems'' that the energy transformation is already realized. 
\par In this paper, the authors emphasize the following opinions about the energy correction based machine transformation
\vspace*{0.5em}
\\ (i) The trajectory transformation completely fails due to the pseudo ECIM.
\\ (ii) The energy transformation is a ``mathematical coincidence'' because it is physically meaningless.
\vspace*{0.5em}
\par Clarifications are given in the following sections.

\subsection{FAILURE OF TRAJECTORY TRANSFORAMTION} \label{section_VD}
\noindent \textit{Statement}: The problem of the pseudo ECIM can be reflected in the failure of the trajectory transformation, because ECIM does not have trajectory that is simulated from the equation of motion.
\\ \textit{Clarification}: The clarification below is given through the only kinetic-energy corrections for simplicity and clearance.
\par Following Eq. (\ref{equ20}) as given in Section \ref{section_VB}, the velocity of the ECIM is given as
\begin{equation}
  \label{equ34}
  \omega_{i\mbox{-}\mathrm{SYS}}^{(\mathrm{EC})}=\sqrt{\frac{2 V_{K E i\mbox{-}\mathrm{SYS}}^{(\mathrm{EC})}}{M_{i}}}=\sqrt{\frac{2\left(V_{K E i\mbox{-}\mathrm{SYS}}-V_{K E i\mbox{-}\mathrm{CR}}\right)}{M_{i}}}
\end{equation}
\par In Eq. (\ref{equ34}), note that $\omega_{i\mbox{-}\mathrm{SYS}}^{(\mathrm{EC})}$ will not exist if IGMKE is larger than IMKE ($V_{K E i\mbox{-}\mathrm{SYS}}^{(\mathrm{EC})}<0$).
\par Following Eqs. (\ref{equ33}) and (\ref{equ34}), one can obtain the following
\begin{equation}
  \label{equ35}
  \omega_{i\mbox{-}\mathrm{SYS}}^{(\mathrm{EC})}<\omega_{i\mbox{-}\mathrm{SYS}}
\end{equation}
\par Eq. (\ref{equ35}) holds only when IGMKE is lower than IMKE.
\par Based on Eq. (\ref{equ35}), in each simulation step, the ECIM trajectory (ECIMTR) is given as
\begin{equation}
  \label{equ36}
  \delta_{i\mbox{-}\mathrm{SYS}, t+1}^{(\mathrm{EC})}=\delta_{i\mbox{-}\mathrm{SYS}, t}^{(\mathrm{EC})}+\omega_{i\mbox{-}\mathrm{SYS}}^{(\mathrm{EC})} \Delta t
\end{equation}
\par From Eqs. (\ref{equ35}) and (\ref{equ36}), it is clear that the followings hold
\begin{equation}
  \label{equ37}
  \delta_{i\mbox{-}\mathrm{SYS}}^{(\mathrm{EC})}<\delta_{i\mbox{-}\mathrm{SYS}}, \quad \delta_{i\mbox{-}\mathrm{SYS}}^{(\mathrm{EC})} \neq \delta_{\mathrm{CR}\mbox{-}\mathrm{SYS}}
\end{equation}
\par From Eqs. (\ref{equ36}) and (\ref{equ37}), $\delta_{i\mbox{-}\mathrm{SYS}}^{(\mathrm{EC})}$ becomes discrete in each simulation step because it is not simulated through the equation of motion.
In fact, $\delta_{i\mbox{-}\mathrm{SYS}}^{(\mathrm{EC})}$ even disappears when IGMKE is higher than IMKE, because ECIMKE becomes negative under this circumstance, as in Fig. \ref{fig19a}.
Therefore, the trajectory transformation completely fails. Detailed analysis about this failure will be given in the case study.

\subsection{MEANINGLESS ENERGY COMPUTATION}  \label{section_VE}
\vspace*{0.5em}
\noindent 1) MEANINGLESS ENERGY CORRECTION UNDER DIFFERENT MOTION REFERENCES
\vspace*{0.5em}
\par Following the analysis in Section \ref{section_VB}, the ECIM is a typical pseudo machine because it is created through the subtraction of the IGMTE from IMTE.
\par Following the analysis of the Newtonian system as given in Section \ref{section_II}, the gravitational field of each ball in the Stanton planet is different although each ball has the same motion reference. Therefore, the energy computation among these balls is physically meaningless because pseudo ball will be created.
\par In fact, the case of the creation of the pseudo ECIM is even worse. if we take a deep insight into the formation of the pseudo ECIM, one can find that the inner-group machine and the individual machine are defined under the different motion references. The definitions of the inner-group machines and individual machines are shown in Figs. \ref{fig21} (a) and (b), respectively.
\begin{figure} [H]
  \centering 
  \subfigure[]{%
  \label{fig21a}
    \includegraphics[width=0.4\textwidth]{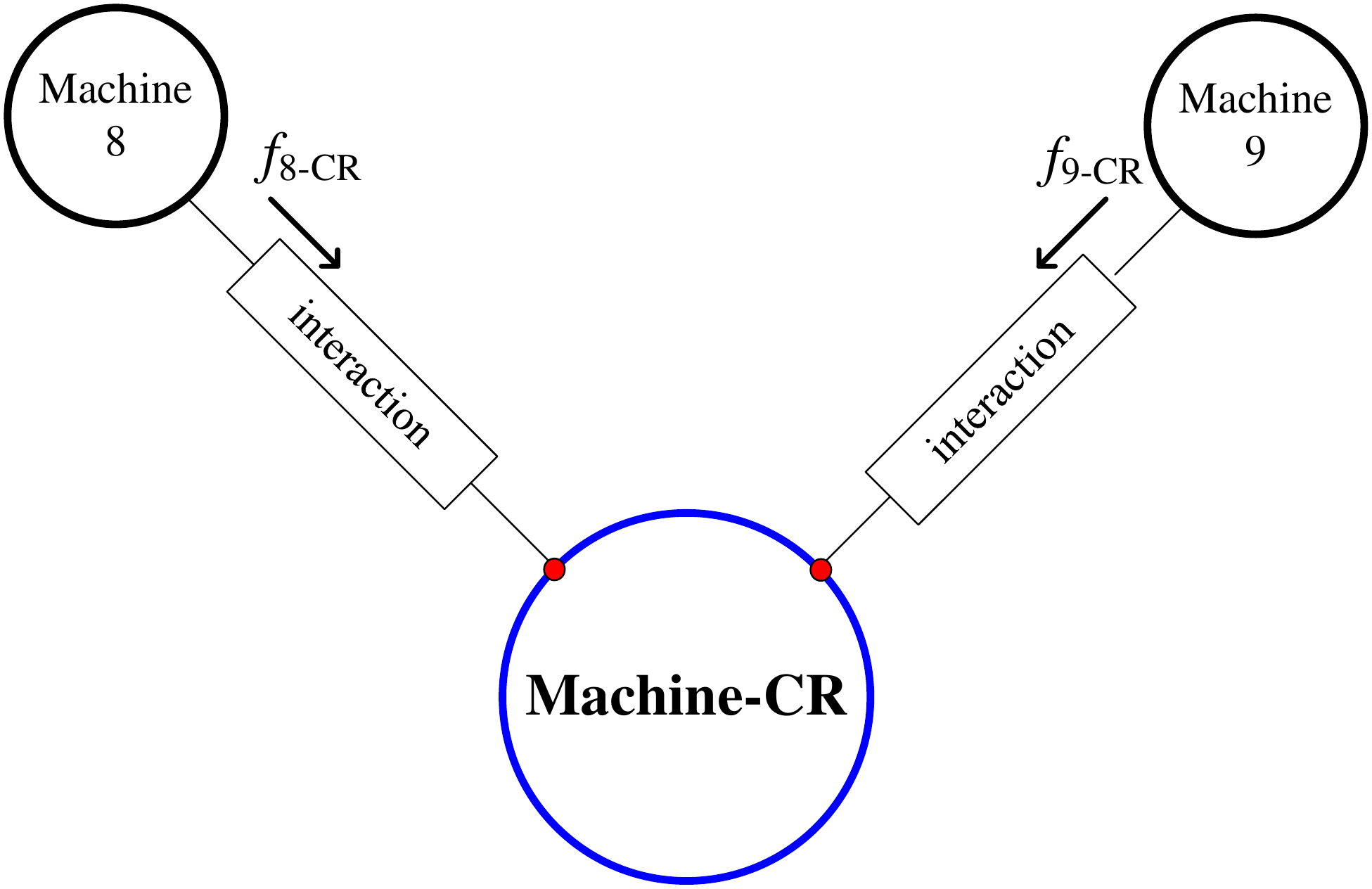}}%
\end{figure} 
\vspace*{-2em}
\addtocounter{figure}{-1}       
\begin{figure} [H]
  \addtocounter{figure}{1}      
  \centering 
  \subfigure[]{%
    \label{fig21b}
    \includegraphics[width=0.4\textwidth]{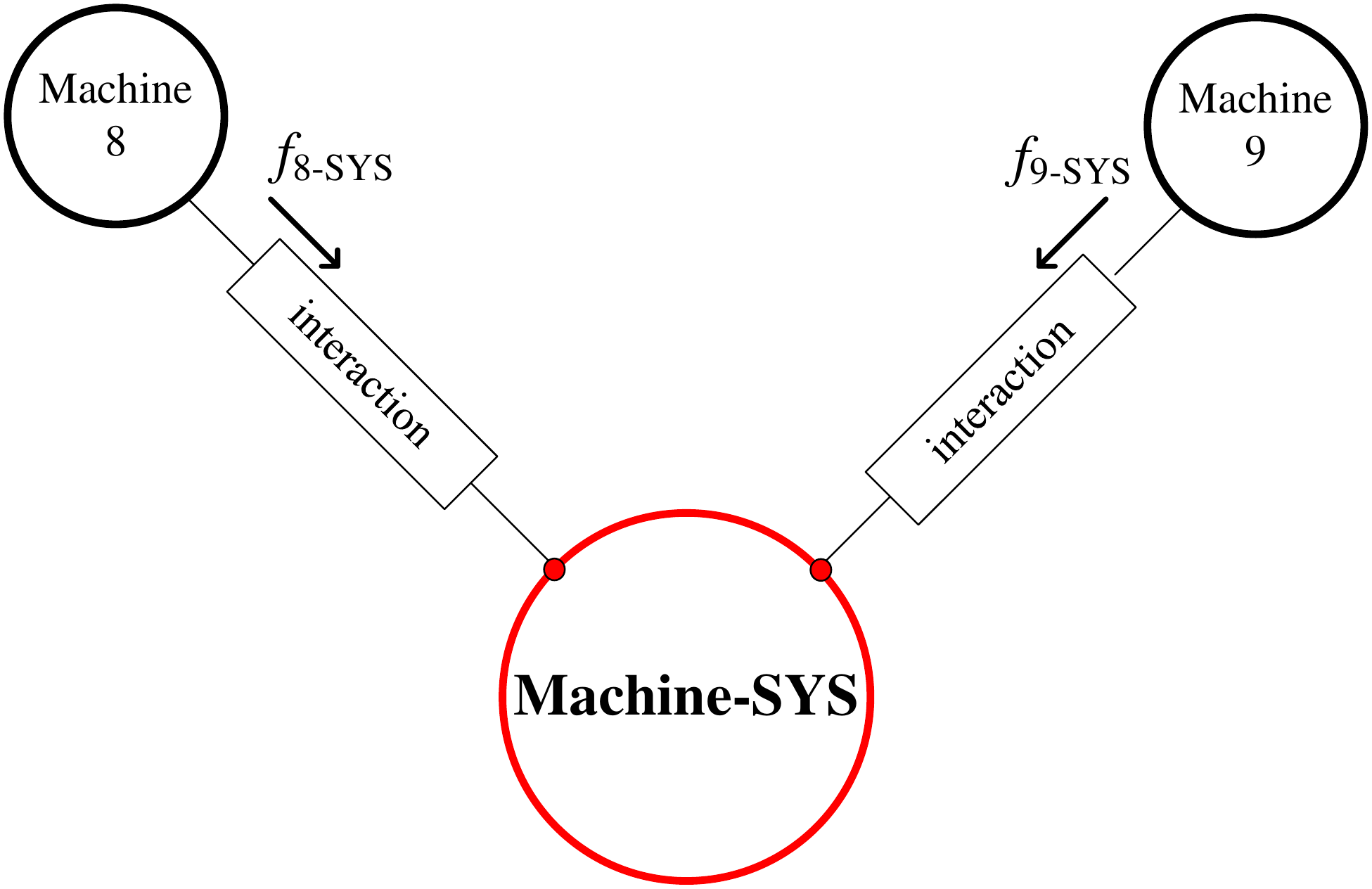}}%
  \caption{The definitions of inner-group machines and individual machines under different motion references. (a) Inner-group machine in the COI-CR reference. (b) Individual machine in the COI-SYS reference.}%
  \label{fig21}
\end{figure}
\vspace*{-0.5em}
\par From Fig. \ref{fig21}, the definitions of the two machines under two different motion references reveal that Machines 8 and 9 are modeled in the two different Newtonian systems. Against this background, the mathematical computation of IGMTE and IMTE are also given under the different motion references.
\par
\begin{figure}[H]
  \centering
  \includegraphics[width=0.45\textwidth,center]{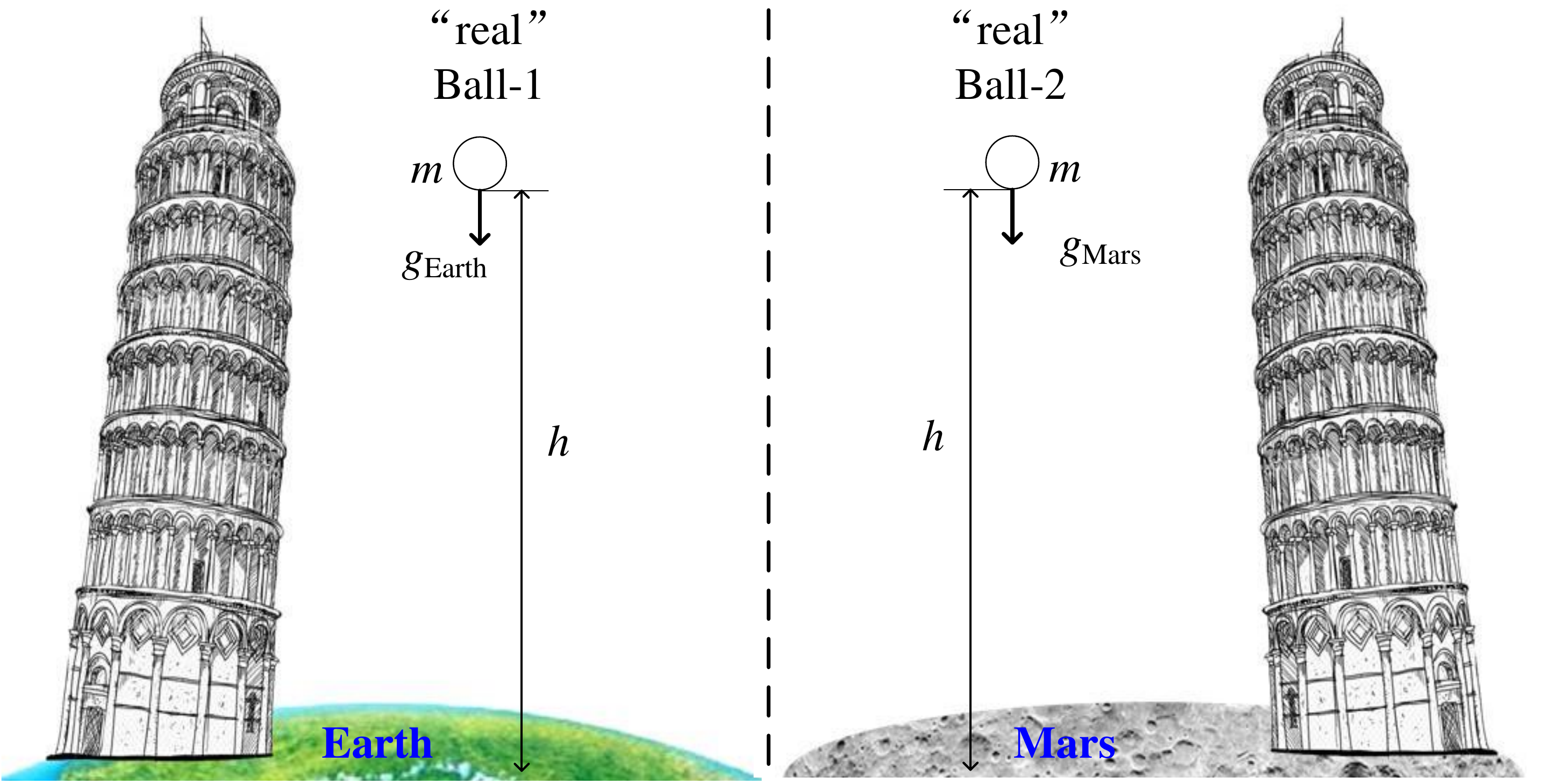}
  \caption{Meaningless energy computation of pseudo balls under different motion reference.} 
  \label{fig22}  
\end{figure}

 This irrationality can be visually explained through comparison between the two balls in the two different planets. The meaningless energy subtraction in the transient energy corrections is shown in Fig. \ref{fig22}.
\vspace*{-0.5em}
\par From Fig. \ref{fig22}, the energy comparisons between the two balls are meaningless because they are defined in the two different Newtonian systems. Based on this, one can simply find that the energy subtraction between IGMTE and IMTE are also physically meaningless because they are defined under the two different motion references, as in Fig. \ref{fig21}.
\vspace*{0.5em}
\\ 2) MEANINGLESS ENERGY SUPERIMPOSITION UNDER THE SAME MOTION REFERENCE
\vspace*{0.5em}
\par
\begin{figure}[H]
  \centering
  \includegraphics[width=0.42\textwidth,center]{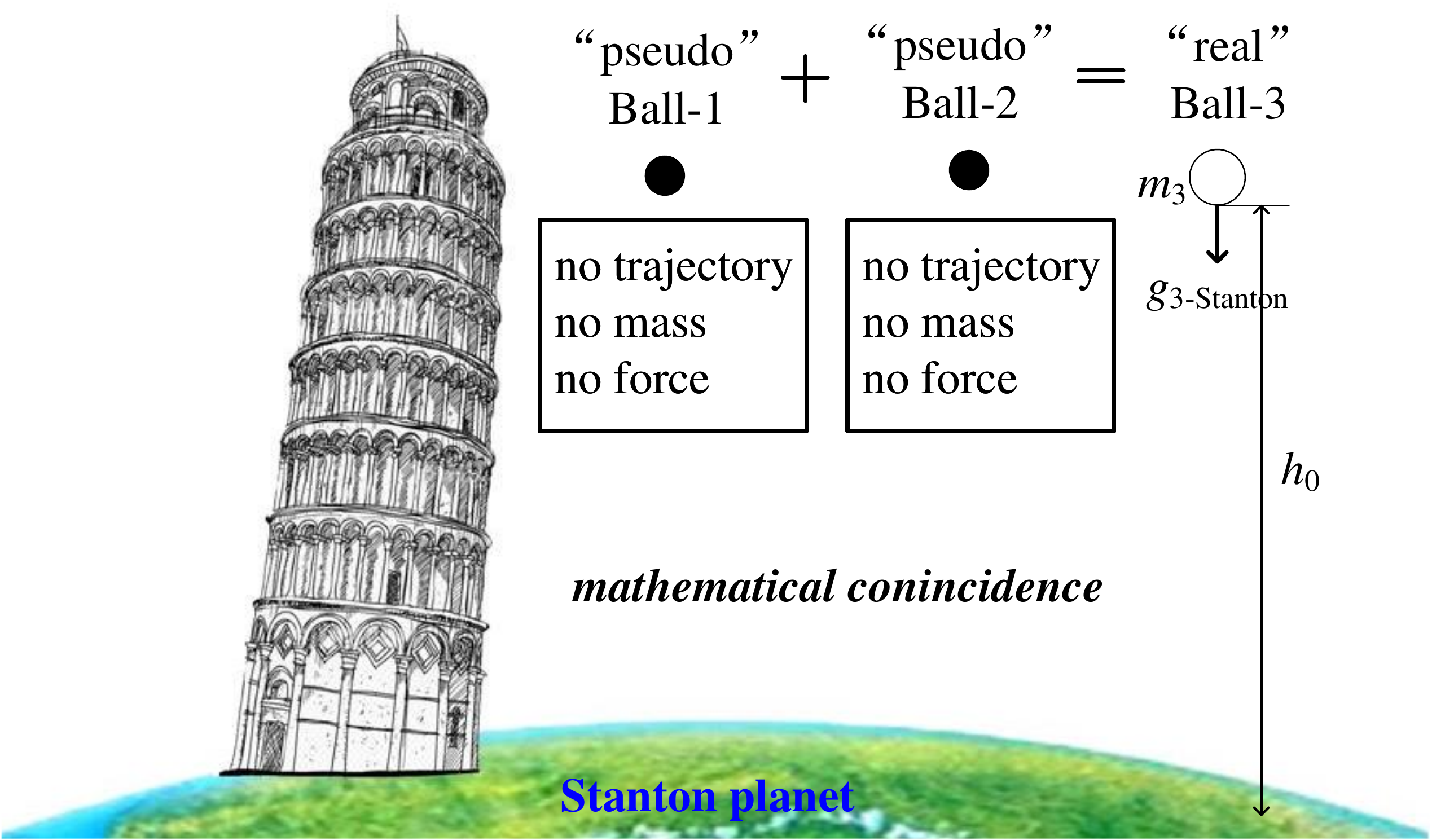}
  \caption{Meaningless energy addition of pseudo balls.} 
  \label{fig23}  
\end{figure}

 Following the analysis in Section \ref{section_VC}, mathematically, the ``real'' equivalent Machine-CR is finally obtained through the superimposition of the ECIMTEs of all the ECIMs inside Group-CR. However, following the analysis in Section \ref{section_II}, this superimposition is physically meaningless because each ECIM is a pseudo machine without equation of motion. This meaningless energy superimposition of the ECIMs is visually shown in Fig. \ref{fig23}.
\vspace*{-0.5em}
\par At this stage, one question naturally emerges: why could the EMTE of real Machine-CR be mathematically obtained through the superimposition of the ECIMTEs, although the energy computation is physically meaningless?
\par In fact, the establishment of $\text{EMTE}_{\text{CR}}$ is a ``mathematical coincidence''. The key reason for this coincidence is given as below
\vspace*{0.5em}
\par \textit{The motions of individual machine, inner-group machine and Machine-CR are not independent}.
\vspace*{0.5em}
\par That is, the motions among these machines have correlations according to their original definitions, as given in Eqs. (\ref{equ25}) and (\ref{equ32}). However, these correlations cannot ensure the flexibility of the machine transformation. This is because the pseudo ECIM is created through the energy subtraction and it does not have trajectory, which causes the trajectory transformation to fail. 
\par In order to solve the problems of the pseudo ECIM, another ``trajectory correction'' based machine transformation is developed. Detailed analysis is given in the following section.

\section{MACHINE TRANSFORMATION THROUGH TRAJECTORY CORRECTIONS}  \label{section_VI}
\subsection{EXPECTED STATE AFTER TRAJECTORY CORRECTIONS} \label{section_VIA}
Following CONJECTURE-II, the action of this machine transformation is given as
\vspace*{0.5em}
\\ Action (trajectory correction): The equation of motion of the inner-group machine is ``corrected'' from the motion of the individual machine.
\vspace*{0.5em}
\par Against this background, the trajectory correction based individual machine (TCIM) is obtained ((the analysis is given in Section \ref{section_VIB}). The expected machine transformations are given as below
\vspace*{0.5em}
\\ Trajectory transformation: The trajectory of TCIM becomes the EMTR.
\\ Energy transformation: the energy transformation is realized after the superimposition of all the TCIM transient energy (TCIMTE) in the system.
\vspace*{0.5em}
\par From the statement above, compared with the energy correction based machine transformation as shown in Figs. \ref{fig17} and \ref{fig18}, the trajectory correction based machine transformation works in a contrary way. In particular, the system engineer firstly focuses on the trajectory correction of the individual machine. Then, this trajectory correction leads to the energy transformation.
\par The expected system state after trajectory corrections are shown in Figs. \ref{fig24} (a) and (b), respectively. The trajectory transformation as given in Fig. \ref{fig18b} is redrawn in Fig. \ref{fig24a}.
\begin{figure} [H]
  \centering 
  \subfigure[]{%
  \label{fig24a}
    \includegraphics[width=0.45\textwidth]{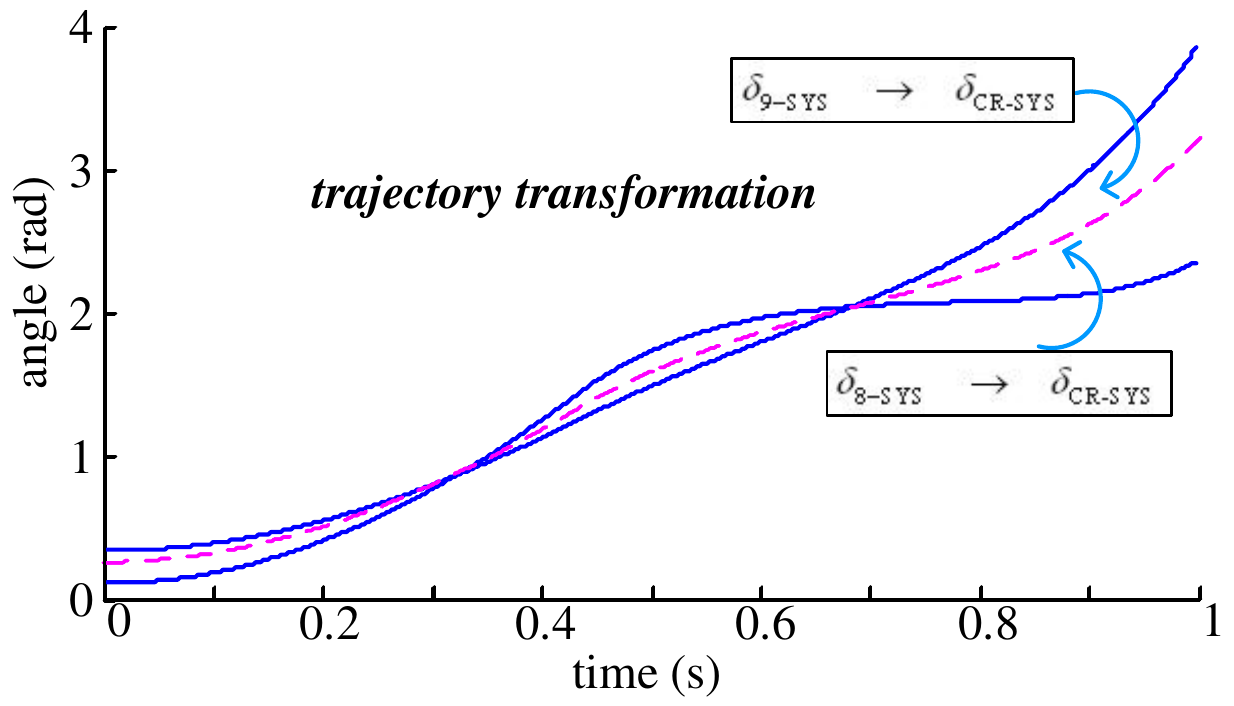}}%
\end{figure} 
\vspace*{-2em}
\addtocounter{figure}{-1}       
\begin{figure} [H]
  \addtocounter{figure}{1}      
  \centering 
  \subfigure[]{%
    \label{fig24b}
    \includegraphics[width=0.45\textwidth]{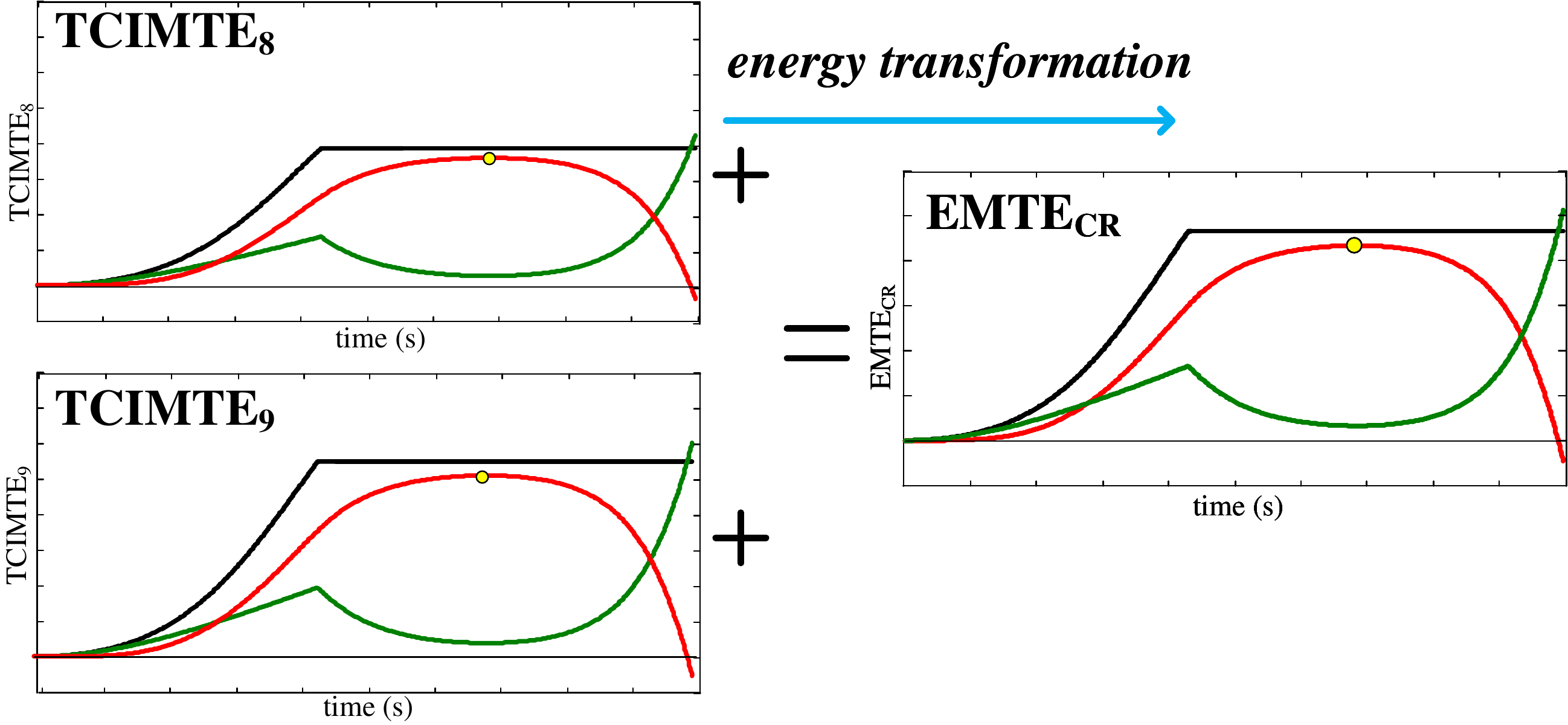}}%
  \caption{Expected system state after trajectory corrections. (a) Trajectory transformation. (b) Energy transformation.}%
  \label{fig24}
\end{figure}

\subsection{SUCCESS OF TRAJECTORY TRANSFORMATION}   \label{section_VIB}
Following Eqs. (\ref{equ12}) and (\ref{equ18}), the trajectory correction of the individual machine is given as
\begin{equation}
  \label{equ38}
  \left\{\begin{array}{l}
    \frac{d\left(\delta_{i\mbox{-}\mathrm{SYS}}-\delta_{i\mbox{-}\mathrm{CR}}\right)}{d t}=\omega_{i\mbox{-}\mathrm{SYS}}-\omega_{i\mbox{-}\mathrm{CR}} \\
   \\
    M_{i} \frac{d\left(\omega_{i\mbox{-}\mathrm{SYS}}-\omega_{i\mbox{-}\mathrm{CR}}\right)}{dt}=f_{i\mbox{-}\mathrm{SYS}}-f_{i\mbox{-}\mathrm{CR}}
    \end{array}\right.
\end{equation}
\par Also based on the expression of $f_{i\mbox{-}\mathrm{CR}}$ as in Eq. (\ref{equ18}), Eq. (\ref{equ38}) can be re-written as
\begin{equation}
  \label{equ39}
  \left\{\begin{array}{l}
    \frac{d \delta_{\mathrm{CR}\mbox{-}\mathrm{SYS}}}{d t}=\omega_{\mathrm{CR}\mbox{-}\mathrm{SYS}} \\
    \\
    M_{i} \frac{d \omega_{\mathrm{CR}\mbox{-}\mathrm{SYS}}}{dt}=\frac{M_{i}}{M_{\mathrm{CR}}} f_{\mathrm{CR}\mbox{-}\mathrm{SYS}}
    \end{array}\right.
\end{equation}
\par In Eq. (\ref{equ39}), the TCIM is obtained through the subtraction of the motion of the inner-group machine from that of the individual machine.
\par  The characteristics of the TCIM are given as below
\vspace*{0.5em}
\\ (i) The equation of motion of each TCIM is replaced with the same motion of Machine-CR, as in Eq. (\ref{equ39}).
\\ (ii) However, the TCIM is different from Machine-CR because its inertia is given as $M_i$ rather than $M_{\text{CR}}$.
\vspace*{0.5em}
\par Following (i) and (ii), the TCIM is also a ``pseudo'' machine because its original motion is destroyed and be replaced with the motion of Machine-CR. Therefore, the TCIM physically does not exist in an actual system.
In fact, the TCIM can be seen as the combination of the individual machine and the equivalent machine. In particular, the original motion of each individual machine is replaced with the motion of Machine-CR, while the inertia of the TCIM is still preserved as $M_i$.
Against this background, the pseudo TCIM still has ``trajectory'' that is the same with that of Machine-CR. The trajectory transformation is finally realized through this trajectory replacement.
\par The trajectory correction can be visually explained through Newtonian mechanics. The mechanisms of the TCIM can be re-depicted as followings
\vspace*{0.5em}
\\ (i) All balls lie under the ``same'' gradational field with the same motion.
\\ (ii) The only difference among these balls is that the inertia ($m_i$) of each ball is different.
\vspace*{0.5em}
\par
\begin{figure}[H]
  \centering
  \includegraphics[width=0.42\textwidth,center]{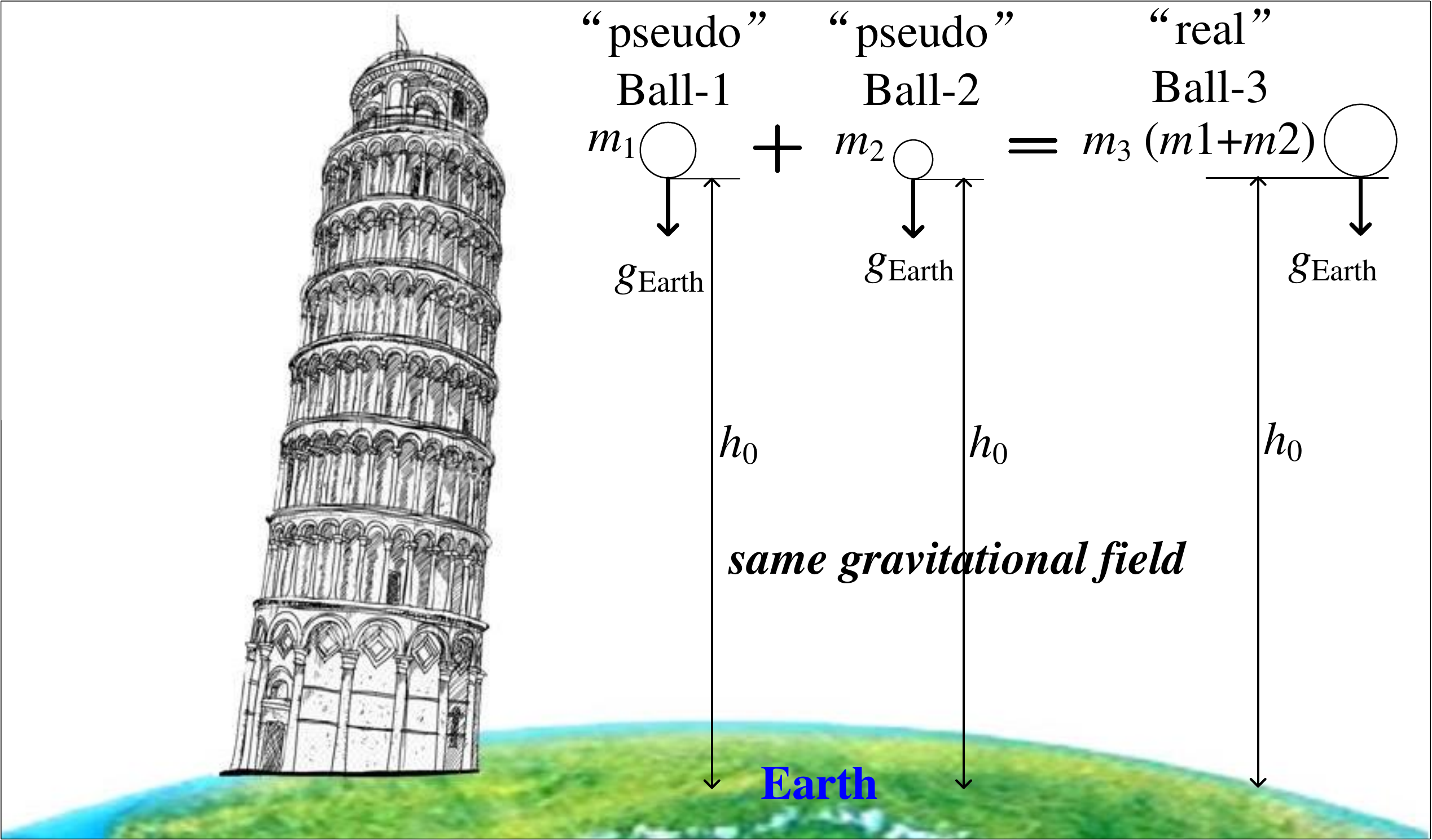}
  \caption{ Explanations of the trajectory corrections through Newtonian mechanics.} 
  \label{fig25}  
\end{figure}

 At this stage, it is clear that (i) and (ii) become the ``big ball-small-ball'' Galileo game as analyzed in Section \ref{section_II}. Explanations of the trajectory corrections using Newtonian mechanics is shown in Fig. \ref{fig25}.
\vspace*{-0.5em}
\par From Fig. \ref{fig25}, since all the balls have the same equation of motion, all these balls lie in the same gravitational field with the same motion reference. Against this background, the energy computation becomes feasible. 

\subsection{SUCCESS OF ENERGY TRANSFORMATION} \label{section_VIC}
Based on Eq. (\ref{equ39}), the TCIM transient energy (TCIMTE) is defined as
\begin{equation}
  \label{equ40}
  V_{i\mbox{-}\mathrm{SYS}}^{(\mathrm{TC})}=V_{K E i\mbox{-}\mathrm{SYS}}^{(\mathrm{TC})}+V_{P E i\mbox{-}\mathrm{SYS}}^{(\mathrm{TC})}
\end{equation}
where
\vspace*{0.4em}
\\
$\left\{\begin{array}{l}
  V_{K E i\mbox{-}\mathrm{SYS}}^{(\mathrm{TC})}=\frac{1}{2} M_{i} \omega_{\mathrm{CR}\mbox{-}\mathrm{SYS}}^{2} \\
  \\
  V_{P E i\mbox{-}\mathrm{SYS}}^{(\mathrm{TC})}=\int_{\delta_{\mathrm{CR}\mbox{-}\mathrm{SYS}}^{s}}^{\delta_{\mathrm{CR}\mbox{-}\mathrm{SYS}}}\left[-\frac{M_{i}}{M_{\mathrm{CR}}} f_{\mathrm{CR}\mbox{-}\mathrm{SYS}}^{(P F)}\right] d \delta_{\mathrm{CR}\mbox{-}\mathrm{SYS}}
  \end{array}\right.$
\vspace*{0.4em}
\par Eq. (\ref{equ40}) can be further depicted as
\begin{equation}
  \label{equ41}
  V_{i\mbox{-}\mathrm{SYS}}^{(\mathrm{TC})}=\frac{M_{i}}{M_{\mathrm{CR}}} V_{\mathrm{CR}\mbox{-}\mathrm{SYS}}
\end{equation}
\par From Eq. (\ref{equ41}), each TCIMTE is the ``scale-down'' ($M_i/M_{\text{CR}}$) of $\text{EMTE}_{\text{CR}}$.
\par Following Eq. (\ref{equ41}), the followings can be obtained
\begin{equation}
  \label{equ42}
  V_{\mathrm{CR}\mbox{-}\mathrm{SYS}}=\sum V_{i\mbox{-}\mathrm{SYS}}^{(\mathrm{MC})}
\end{equation}
\par From Eq. (\ref{equ42}), the energy transformation is successfully established through the energy superimposition of all the TCIMTEs inside Group-CR. In particular, the transient energy of ``real'' Machine-CR is obtained through the superimposition of all the TCIMTEs inside Group-CR.
\par After trajectory correction of each individual machine, the TCIMTE will also become different from IMTE. This difference is depicted as 
\begin{equation}
  \label{equ43}
  \Delta V_{i\mbox{-}\mathrm{SYS}}=\Delta V_{K E i\mbox{-}\mathrm{SYS}}+\Delta V_{P E i\mbox{-}\mathrm{SYS}}
\end{equation}
where
\begin{spacing}{2}
  \noindent$\Delta V_{K E i\mbox{-}\mathrm{SYS}}=\frac{1}{2} M_{i}\left(\omega_{i\mbox{-}\mathrm{SYS}}^{2}-\omega_{\mathrm{CR}\mbox{-}\mathrm{SYS}}^{2}\right)$\\
  $\Delta V_{P E i\mbox{-}\mathrm{SYS}}=\int_{\delta_{i\mbox{-}\mathrm{SYS}}^{s}}^{\delta_{i\mbox{-}\mathrm{SYS}}}\left[-f_{i\mbox{-}\mathrm{SYS}}^{(P F)}\right] d \delta_{i\mbox{-}\mathrm{SYS}}\\
  -\int_{\delta_{\mathrm{CR}\mbox{-}\mathrm{SYS}}^{s}}^{\delta_{\mathrm{CR}\mbox{-}\mathrm{SYS}}}\left[-\frac{M_{i}}{M_{\mathrm{CR}}} f_{\mathrm{CR}\mbox{-}\mathrm{SYS}}^{(P F)}\right] d \delta_{\mathrm{CR}\mbox{-}\mathrm{SYS}}$
\end{spacing}
\par From Eq. (\ref{equ43}), at first glance, it ``seems'' that the trajectory correction is also another form of ``energy correction'' because the energy is changed (the TCIMTE can be obtained through the subtraction of $\Delta V_{i\mbox{-}\mathrm{SYS}}$ from IMTE).
However, this is a misunderstanding. The reason is that the $\Delta V_{i\mbox{-}\mathrm{SYS}}$ is certain to occur after trajectory correction.
In other words, $\Delta V_{i\mbox{-}\mathrm{SYS}}$ is the reflection of the the trajectory correction of each individual machine in a transient energy manner. This is different from the mechanism of the energy correction.

\subsection{FURTHER ANALYSIS ABOUT THE TCIM}  \label{section_VID}
The analysis in Sections \ref{section_V} and this section leads to the emergence of one question: why the energy correction based machine transformation fails while the trajectory correction based machine transformation succeeds?
\par The answer is given as
\vspace*{0.5em}
\par \textit{Transient energy is only used to measure the trajectory (motion) of the machine}.
\vspace*{0.5em}
\par Frankly, the transient energy is a ``tool'' to measure the trajectory stability of the machine. Without any consideration of the trajectory, the energy computation will become physically meaningless. In particular, following the analysis in Section \ref{section_V}, ECIM is created through the subtraction of the IGMTE of the ``real'' inner-group machine through IMTE of the ``real'' individual machine. However, this ``real-minus-real'' creates the pseudo ECIM without equation of motion. Then, the superimposition of all the ECIMTEs inside Group-CR becomes meaningless.
\par Comparatively, the trajectory correction is very distinctive. This is because the trajectory correction can be seen as the ``motion replacement'' of the individual machine with the same equation of motion of Machine-CR.
Through this motion replacement, although the original motion of each individual machine is destroyed and TCIM becomes pseudo, each TCIM has the same motion after trajectory correction, i.e., the equation of motion of Machine-CR. Against this background, each TCIM can be seen as lying in the ``same'' gravitational field, and each TCIMTE becomes the ``scale-down'' of the same $\text{EMTE}_{\text{CR}}$, as in Fig. \ref{fig24}.
The energy superimposition of all the TCIMTE inside Group-CR also becomes flexible. 
\par Comparison between ECIM and TCIM is shown in Table \ref{table2}.
\begin{table}[H]
  \caption{Comparison between ECIM and TCIM.}
  \label{table2}
  \centering
  \begin{tabular}{@{}c|cc@{}}
  \toprule
                                                                      & ECIM                                                           & TCIM                                                                                 \\ \midrule
  Machine type                                                        & pseudo                                                         & pseudo                                                                               \\
  Modeling                                                            & energy correction                                              & \begin{tabular}[c]{@{}c@{}}trajectory correction\\ (motion replacement)\end{tabular} \\
  \begin{tabular}[c]{@{}c@{}}equation of \\ motion\end{tabular}       & N/A                                                            & \begin{tabular}[c]{@{}c@{}}replaced with\\ Machine-CR\end{tabular}                   \\
  inertia                                                             & N/A                                                            & $M_i$                                                                                   \\
  NEC                                                                 & N/A                                                            & \begin{tabular}[c]{@{}c@{}}``scale-down'' of\\ Machine-CR\end{tabular}               \\
  \begin{tabular}[c]{@{}c@{}}Trajectory\\ transformation\end{tabular} & fail                                                           & success                                                                              \\
  \begin{tabular}[c]{@{}c@{}}Energy\\ transformation\end{tabular}     & \begin{tabular}[c]{@{}c@{}}mathematically\\ holds\end{tabular} & success                                                                              \\ \bottomrule
  \end{tabular}
\end{table}
\vspace*{-1em}
\par From Table \ref{table2}, in brief, the failure of trajectory transformation in the ECIM is caused by the missing of equation of motion. Comparatively, the successes of both trajectory transformation and energy transformation in the TCIM are fully ensured by the motion replacement of Machine-CR. This further validates that transient energy is only used to measure the trajectory of the machine. That is, the energy computations among machines are generally meaningless, unless the motion of each machine is replaced with the same motion.

\section{CASE STUDY} \label{section_VII}
\subsection{FAILURE OF TRAJECTORY TRANSFORMATION THROUGH ENERGY CORRECTIONS} \label{section_VIIA}
\begin{figure}[H]
  \centering
  \includegraphics[width=0.42\textwidth,center]{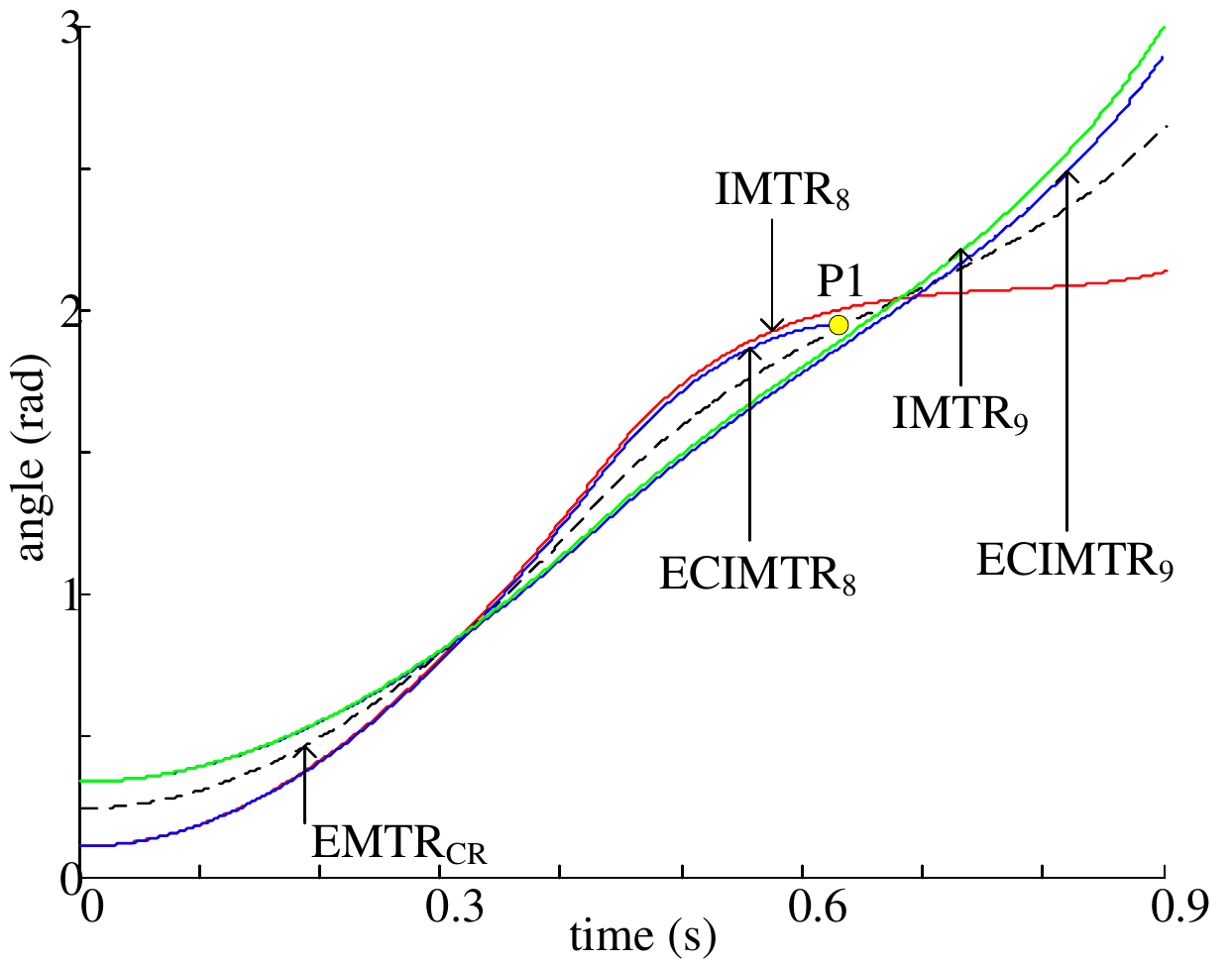}
  \caption{Failure of trajectory transformation.} 
  \label{fig26}  
\end{figure}

The failure of the trajectory transformation of the ECIM is demonstrated through the case [TS-1, bus-2, 0.430 s]. The trajectory transformation of ECIM is given in Fig. \ref{fig26}. Note that trajectory transformation is based on the only kinetic energy correction as analyzed in Section Section \ref{section_VE}. The ``success'' of energy transformation through energy corrections is already analyzed in Section \ref{section_VC}.
\vspace*{-0.5em}
\par From Fig. \ref{fig26}, after energy correction of IGMKE as given in Fig. \ref{fig16}, each ECIMKE becomes lower than IMKE. Therefore, the velocity of the ECIM is smaller than that of corresponding individual machine, as analyzed in Section \ref{section_VD}.
Against this background, each ECIMTR moves ``downward'' compared with IMTR, as in Fig. \ref{fig26}. Especially, for the case of Machine 8, because the $\text{ECIMKE}_8$ becomes negative at P1 (0.627 s), the $\text{ECIMTR}_8$ even becomes ``broken'' along time horizon.
\par From analysis above, each ECIMTR is far from the expected $\text{EMTR}_{\text{CR}}$. This fully indicates that trajectory transformation completely fails in the ECIM.

\subsection{SUCCESS OF ENERGY TRANSFORMATION THROUGH TRAJECTORY CORRECTIONS}  \label{section_VIIB}
Compared with the case in Section \ref{section_VIIA}, each TCIMTR becomes the $\text{EMTR}_{\text{CR}}$ after trajectory correction, because the equation of motion of each TCIM is replaced with that of Machine-CR, as analyzed in Section \ref{section_VIA}. The TCIMTE is shown in Figs. \ref{fig27} (a) and (b), respectively. Note that $\text{EMTE}_{\text{CR}}$ is already shown in Fig. \ref{fig14}.
\begin{figure} [H]
  \centering 
  \subfigure[]{%
  \label{fig27a}
    \includegraphics[width=0.45\textwidth]{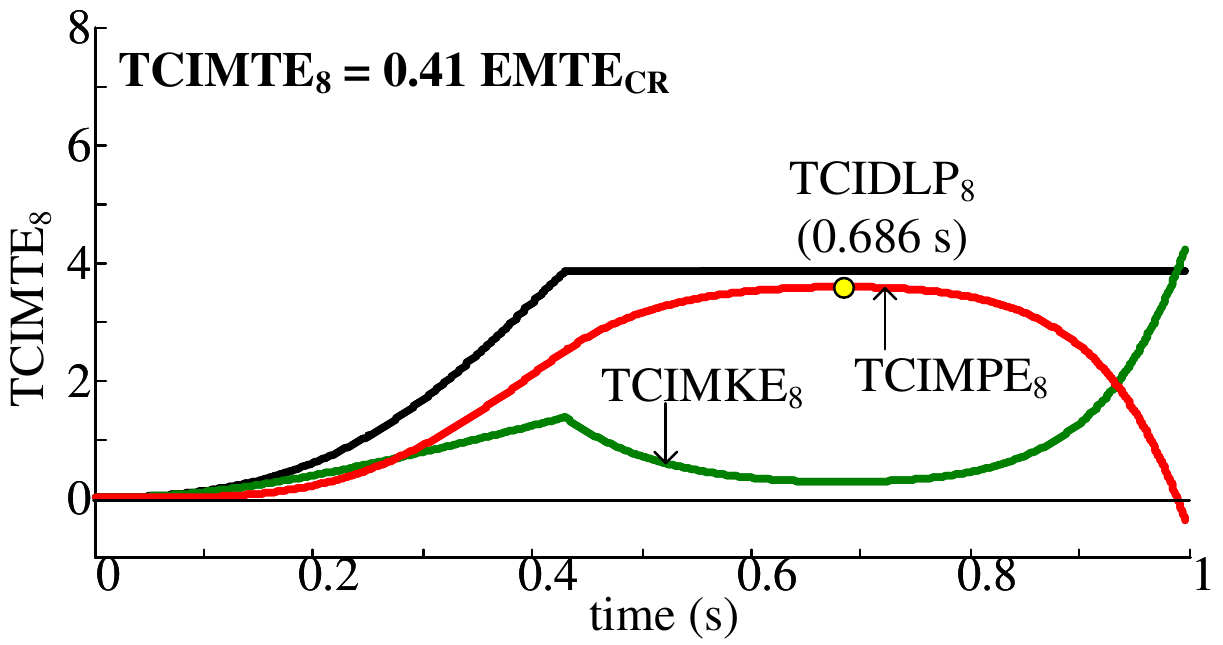}}%
\end{figure} 
\vspace*{-2em}
\addtocounter{figure}{-1}       
\begin{figure} [H]
  \addtocounter{figure}{1}      
  \centering 
  \subfigure[]{%
    \label{fig27b}
    \includegraphics[width=0.45\textwidth]{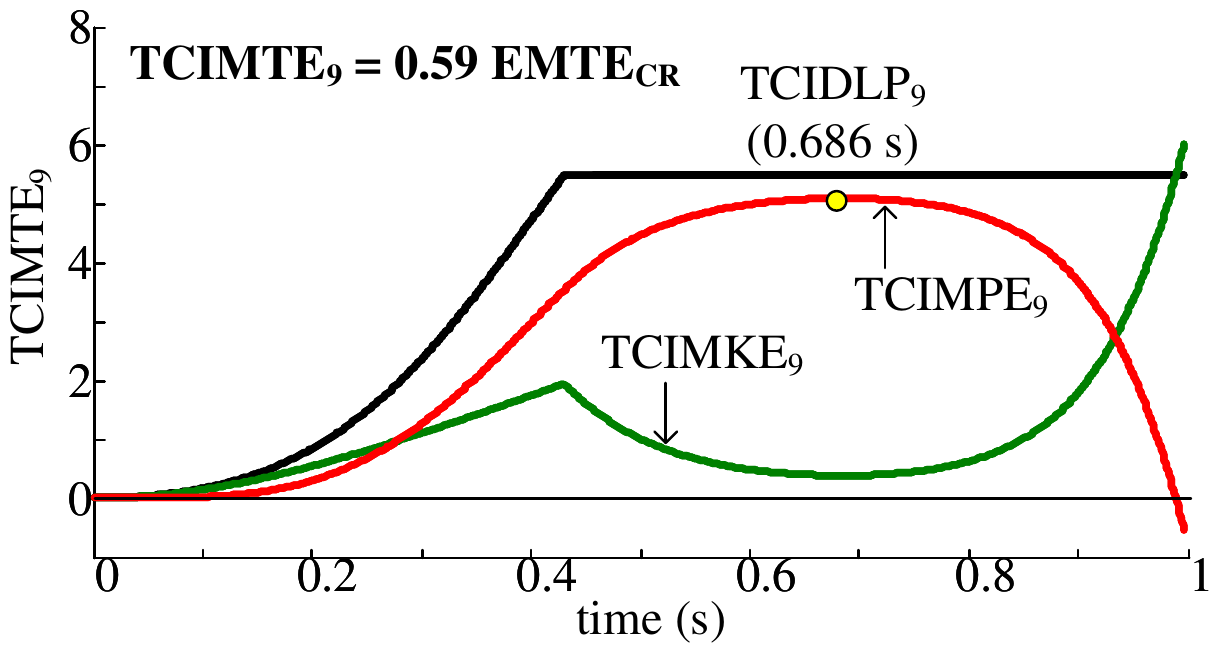}}%
  \caption{TCIMTE after trajectory corrections. (a) $\text{TCIMTE}_8$. (b) $\text{TCIMTE}_9$.}%
  \label{fig27}
\end{figure}
\vspace*{-0.5em}
\par From Fig. \ref{fig26}, $\text{TCIMTE}_8$ and $\text{TCIMTE}_9$ are the ``scale down'' of $\text{EMTE}_{\text{CR}}$. The superimposition of them is just equal to $\text{EMTE}_{\text{CR}}$.
In addition, $\text{TCIDLP}_8$ and $\text{TCIDLP}_9$ occur simultaneously. This is because the motions of $\text{TCIM}_8$ and $\text{TCIM}_9$ are completely the same with that of Machine-CR. This fully indicates that the energy transformation is realized after the trajectory correction.
\par This successful energy transformation can also be demonstrated from the perspective of $\Delta V_{i\mbox{-}\mathrm{SYS}}$. $\Delta V_{\mathrm{KE8}\mbox{-}\mathrm{SYS}}$
and $\Delta V_{\mathrm{KE9}\mbox{-}\mathrm{SYS}}$ along time horizon are shown in Figs. \ref{fig28} (a) and (b), respectively.
The velocities of the two machines and Machine-CR are shown in Fig. \ref{fig29}.
\begin{figure} [H]
  \centering 
  \subfigure[]{%
  \label{fig28a}
    \includegraphics[width=0.45\textwidth]{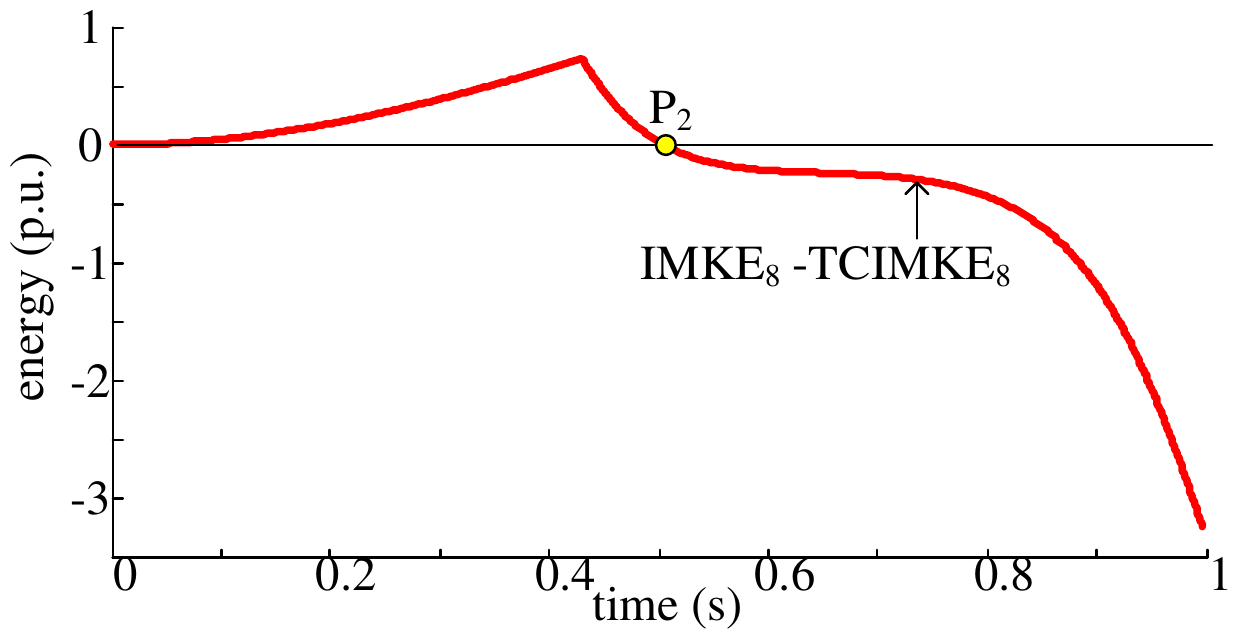}}%
\end{figure} 
\vspace*{-2em}
\addtocounter{figure}{-1}       
\begin{figure} [H]
  \addtocounter{figure}{1}      
  \centering 
  \subfigure[]{%
    \label{fig28b}
    \includegraphics[width=0.45\textwidth]{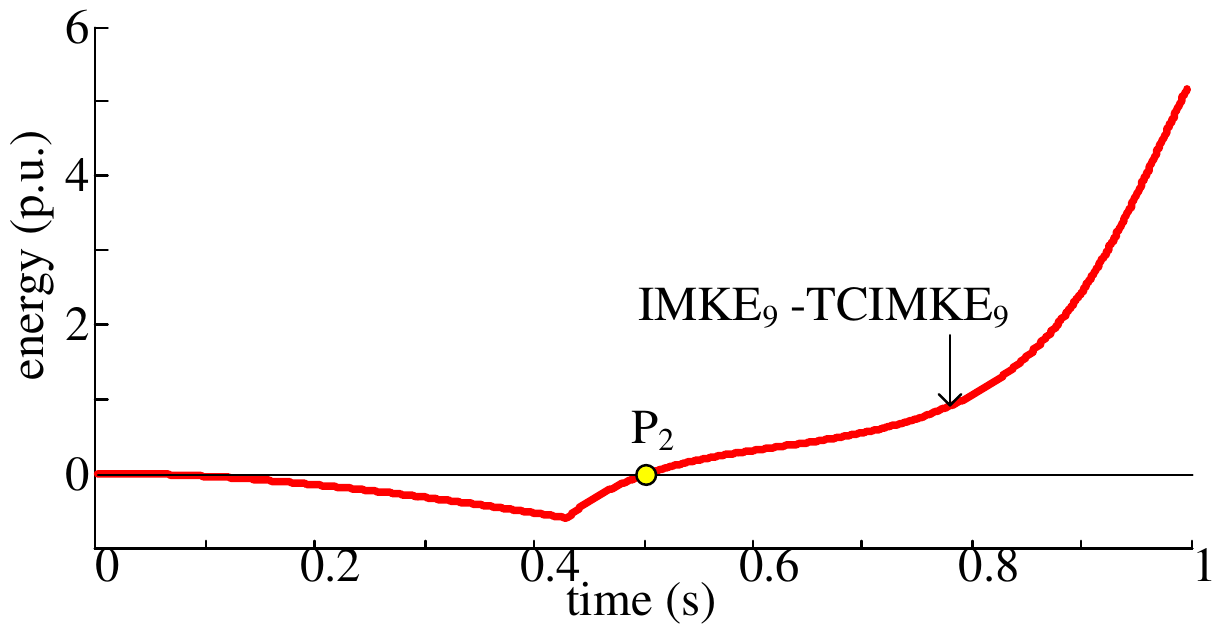}}%
  \caption{Demonstration of energy variance of the individual machine after trajectory corrections. (a) $\Delta V_{\mathrm{KE8}\mbox{-}\mathrm{SYS}}$. (b) $\Delta V_{\mathrm{KE9}\mbox{-}\mathrm{SYS}}$.}%
  \label{fig28}
\end{figure}
\vspace*{-1em}
\begin{figure}[H]
  \centering
  \includegraphics[width=0.45\textwidth,center]{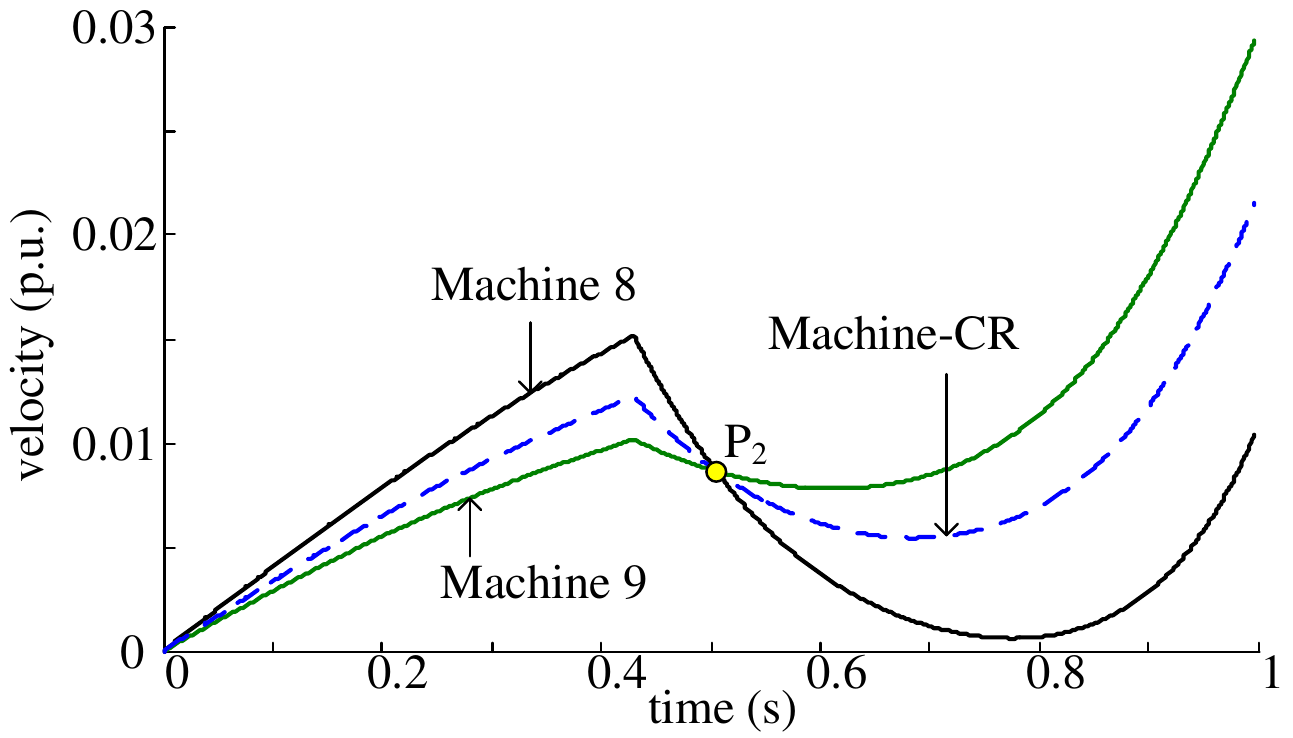}
  \caption{Velocities of the individual machines and Machine-CR.} 
  \label{fig29}  
\end{figure}
\vspace*{-0.5em}
For the case in Fig. \ref{fig27}, because Machines 8 and 9 form Group-CR, and thus the following holds
\begin{equation}
  \label{equ44}
  M_{8} \omega_{8\mbox{-}\mathrm{SYS}}+M_{9} \omega_{9\mbox{-}\mathrm{SYS}}=M_{\mathrm{CR}} \omega_{\mathrm{CR}\mbox{-}\mathrm{SYS}}
\end{equation}
\par In Fig. \ref{fig27}, P2 is the point where $\omega_{8\mbox{-}\mathrm{SYS}}$ is equal to $\omega_{9\mbox{-}\mathrm{SYS}}$. Following Eq. (\ref{equ44}), $ \omega_{\mathrm{CR}\mbox{-}\mathrm{SYS}}$
is also equal to $\omega_{8\mbox{-}\mathrm{SYS}}$ and $\omega_{9\mbox{-}\mathrm{SYS}}$. Therefore, both $\Delta V_{\mathrm{KE}8\mbox{-}\mathrm{SYS}}$ and $\Delta V_{\mathrm{KE}9\mbox{-}\mathrm{SYS}}$ reach zero simultaneously at P2.
\par Taking the period from the pre-fault point to P2 as an example, from Fig. \ref{fig28}, $\Delta V_{\mathrm{KE}8\mbox{-}\mathrm{SYS}}$ and $\Delta V_{\mathrm{KE}9\mbox{-}\mathrm{SYS}}$ are positive and negative, respectively.
This is because $\omega_{8\mbox{-}\mathrm{SYS}}$ is larger than $\omega_{\mathrm{CR}\mbox{-}\mathrm{SYS}}$, and $\omega_{9\mbox{-}\mathrm{SYS}}$ is smaller than $\omega_{\mathrm{CR}\mbox{-}\mathrm{SYS}}$ during this period.
Therefore, $\Delta V_{\mathrm{KE}8\mbox{-}\mathrm{SYS}}$ and $\Delta V_{\mathrm{KE}9\mbox{-}\mathrm{SYS}}$ are caused by trajectory corrections of the two machines. They are also the reflections of the trajectory corrections of the two machines in a transient energy manner.
This is different from the energy correction case in which both $\text{IGMKE}_8$ and $\text{IGMKE}_9$ (the kinetic energy that is corrected from the IMKE) are positive, as shown in Fig. \ref{fig16}.
\par From simulations above, only trajectory corrections, i.e., the motion replacement in each individual machine may leads to the success of machine transformation, as analyzed in Section \ref{section_VI}.

\section{CONCLUSIONS} \label{section_VIII}
In this paper, the transformations from the individual machine to the equivalent machine are analyzed through the ``energy correction'' and ``trajectory correction'' of the inner-group machine. The balls in the Newtonian system are first classified as the real ball and pseudo ball. The ball classifications of balls are naturally extended to the machine classifications in the power system transient stability. It is clarified that both individual machine and equivalent machine are real machines with equation of motions.
Based on the machine classifications, two types of machine transformations are analyzed. For the energy correction case, it is found that the ECIM is a pseudo machine without equation of motion and thus the trajectory transformation completely fails. Meanwhile, the superimposition of the ECIMTEs is a mathematical coincidence because this superimposition is physically meaningless. For the trajectory correction case, the TCIM is also a pseudo machine. 
However, each TCIM has the same equation of motion, i.e., the motion of Machine-CR after motion replacement. Against this background, both trajectory transformation and the energy transformation are finally realized by using TCIM. Simulation results show that the energy correction may cause the ECIMTR to move downward in the original system, and this indicates that the trajectory transformation completely fails. Comparatively, the trajectory fully ensures the flexibility of both energy transformation and trajectory transformation. In addition, it is clarified that the change of TCIMTE compared with IMTE are caused by trajectory corrections of the each machine. The change in TCIMTE the reflections of the trajectory corrections of the two machines in a transient energy manner.
\par The studies in this paper help readers take a deep insight into the physical nature of the transient energy. Most importantly, the complicated relationships among individual machine, equivalent machine, inner-group machine and the superimposed machine are systematically clarified through another distinctive perspective of machine transformation rather than the original definitions.

%

%
%
%




\end{document}